\documentclass{aastex631}

\usepackage{longtable}
\usepackage{graphicx} 
\usepackage{colortbl,booktabs}
\usepackage{indentfirst}
\usepackage{gensymb} 
\usepackage{natbib}
\usepackage{subfigure}
\usepackage{appendix}
\usepackage{makecell,rotating,multirow,diagbox}
\usepackage{threeparttable}
\usepackage{amssymb}
\usepackage{pifont}

\newcommand{\lya}{Ly$\alpha$}
\newcommand{\alfven}{Alfv\'{e}n}

\graphicspath{{./}{figures/}}

\begin{document}
	
\title{The Lyman-alpha Emission in Solar Flares. II. A Statistical Study on Its Relationship with the White-light plus Soft X-ray Emission}
	
\author[0000-0003-0057-6766]{De-Chao Song}
\affiliation{Key Laboratory of Dark Matter and Space Astronomy, Purple Mountain Observatory, CAS, Nanjing 210023, People's Republic of China}
	
\author[0000-0002-8258-4892]{Ying Li}
\affiliation{Key Laboratory of Dark Matter and Space Astronomy, Purple Mountain Observatory, CAS, Nanjing 210023, People's Republic of China}
\affiliation{School of Astronomy and Space Science, University of Science and Technology of China, Hefei 230026, People's Republic of China}
	
\author[0000-0001-7540-9335]{Qiao Li}
\affiliation{Key Laboratory of Dark Matter and Space Astronomy, Purple Mountain Observatory, CAS, Nanjing 210023, People's Republic of China}
		
\author[0000-0002-3657-3172]{Xiaofeng Liu}
\affiliation{Key Laboratory of Dark Matter and Space Astronomy, Purple Mountain Observatory, CAS, Nanjing 210023, People's Republic of China}
\affiliation{School of Astronomy and Space Science, University of Science and Technology of China, Hefei 230026, People's Republic of China}

\correspondingauthor{De-Chao Song and Ying Li}
\email{dcsong@pmo.ac.cn, yingli@pmo.ac.cn}

\begin{abstract}
The hydrogen \lya\ line and the white-light (WL) continuum are two key diagnostics of energy transport in the lower atmosphere during solar flares, yet their relationship remains poorly understood. Here we present a statistical analysis of 69 white-light flares (WLFs) to investigate the relationships among the \lya, soft X-ray (SXR), and WL continuum emissions using the data from GOES and the Helioseismic and Magnetic Imager (HMI) on the Solar Dynamics Observatory. We find that the \lya\ contrast in these WLFs ranges 0.8--28.5\% with a mean value of 7.0\%. Positive power-law relationships exist among peak enhancements in SXR, \lya, and WL. For most events, the \lya\ peak is nearly co-temporal with the peak of SXR time derivative, whereas the WL peak is either co-temporal with or lags those of \lya\ and SXR derivative. The \lya\ and WL rise times are similar ($\sim$3--4 min) and correlated. We also find that the radiated energy in \lya\ and HMI narrow-band WL has a positive power-law relationship with duration. In particular, the power-law index for the narrow-band WL is very close to 1/3 as predicted by magnetic reconnection theory. On average, the radiated energies in GOES \lya\ and SXR bands are approximately three orders of magnitude greater than the energy emitted in the continuum near 6173 \AA\ with a bandwidth of 1 \AA. Our findings provide new constraints on lower-atmosphere energy transport in solar flares and can serve as valuable references for modelling and interpreting the flares on solar-type stars.
\end{abstract}
	
\keywords{Solar activity (1475); Solar flares (1496); Solar white-light flares (1983); Solar ultraviolet emission (1533)}

\section{Introduction}
\label{intro}

Solar flares are among the most violent explosive phenomena in the solar system, producing enhanced radiation across nearly the entire electromagnetic spectrum, from $\gamma$-rays to radio waves. In many large flares, a significant fraction of the total radiated energy originates from the ultraviolet and white-light (WL, i.e., visible) bands. The hydrogen \lya\ (\ion{H}{1} \lya) line at 1216 \AA\ and the WL continuum (3800--7600 \AA; \citealp{Tobiska2006}) are two key spectral windows for characterizing these flare emissions.

The \lya\ line, originating from the 2p--1s transition of neutral hydrogen, is the strongest line in the solar vacuum ultraviolet spectrum \citep{Curdt2001AA}. It is formed primarily in the middle-to-upper chromosphere and the lower transition region \citep{Vernazza1981}. During flares, both the line core and wings exhibit varying degrees of enhancement, which has been confirmed by theoretical calculations \citep[e.g.,][]{Henoux1995,Zhao1998}, spectroscopic observations \citep[e.g.,][]{Canfield1980,Brekke1996,Woods2004}, and radiative hydrodynamic simulations \citep[e.g.,][]{Allred2005,Hong2019,Yyt2021RAA,Tian2022}. To date, research on flare \lya\ emission has predominantly relied on irradiance observations \citep[e.g.,][]{Johnson2011,Milligan2012,Kretzschmar2013,Milligan2016,Milligan2017,Chamberlin2018,Dominique2018,LiD2020,Jing2020,Lu2021catalog,Lu2021QPP,Milligan2021,Tian2023,Greatorex2023,Majury2025}. These data come from instruments such as the Extreme Ultraviolet Sensor \citep[EUVS;][]{Evans2010} onboard the Geostationary Operational Environmental Satellite (GOES), the Extreme ultraviolet Variabilty Experiment \citep[EVE;][]{Woods2012} on the Solar Dynamics Observatory \citep[SDO;][]{Pesnell2012}, and the Large Yield Radiometer \citep[LYRA;][]{Dominique2013} on the Project for On-Board Autonomy 2 (PROBA2). Statistical analyses show that the relative enhancement in flare \lya\ irradiance is typically below 30\%, with the majority of flares showing an increase of $\le$ 10\% \citep[e.g.,][]{NusinovKaza2006,Raulin2013,Milligan2020,Lu2021catalog}. The relatively rare \lya\ imaging studies suggest that the enhancement arises mainly from flare footpoints and is closely associated with the heating of the lower atmosphere by nonthermal electron bombardment \citep[e.g.,][]{RubiodaCosta2009,LiD2024}, implying a nonthermal origin.

Flare \lya\ emission can also have a thermal origin, particularly in smaller B- and C-class flares. For instance, \citet{Milligan2021} conducted a superposed-epoch analysis of 3123 B-class and 4972 C-class flares. Their results showed that for B-class flares, the average \lya\ peak time coincides with the soft X-ray (SXR) peak, whereas for C-class flares, the \lya\ peak slightly precedes the SXR peak. Recent imaging observations lend further support to a thermal origin in small flares. \citet{LiY2022}, for example, analyzed \lya\ and hard X-ray (HXR) observations from the Solar Orbiter \citep[SolO;][]{SO2020} for a C1.4 flare, finding a strong temporal correlation between the \lya\ emission and the thermal emission observed in SXR 1--8 \AA\ and HXR at 5--7 keV. Furthermore, a thermal origin for \lya\ emission might be dominant in some M- and X-class flares. In the first paper of this series \cite[][]{Jing2020}, a study of more than 650 M- and X-class flares observed by GOES during 2006--2016 showed that 143 events (21.7\%) exhibit a \lya\ peak lagging the SXR 1--8 \AA\ peak. This lag is potentially attributable to the cooling of thermal plasma.

Flares with significant enhancement of WL continuum are termed white-light flares \citep[WLFs;][]{Fletcher2011,Hudson2016}. The WL continuum enhancement can arise from free-bound (recombination) radiation in the chromosphere and/or from increased $H^{-}$ emission in the photosphere, depending on the depth of energy deposition \citep[e.g.,][]{Machado1978,Emslie1982,Gan1992,Fang1995AAS,Ding2003,Xu2006,Fletcher2008,Heinzel2014,Watanabe2020,Xu2025,Tian2025}. WLFs are detected most often in M- and X-class events \citep[e.g.,][]{Castellanos2020,LiY2024a,LiY2024b,JZC2024,JZC2025}, with fewer reports in C-class flares \citep[e.g.,][]{Hudson2006,Song2018jj,Song2018aa,Song2020,LiQ2024,Cai2024,Xu2025}. The origin of WL continuum emission is closely tied to direct and indirect heating of the chromosphere and even a deeper layer. 
Observationally, WL continuum enhancements show a close spatiotemporal correlation with HXR (and sometimes $\gamma$-ray) emission \citep[e.g.,][]{Kuhar2016,SDC2023,Song2025,Battaglia2025,LiD2025}. This behavior is similar to that of \lya\ emission during the impulsive phase of most flares, suggesting a common driver such as magnetic reconnection. However, the relationship between \lya\ and WL continuum during flares remains poorly understood.

In this paper, as the second one of the series, we present a statistical analysis of the relationship between \lya\ and WL continuum emission properties in 69 WLFs. The analysis combines \lya\ irradiance data from GOES with WL continuum intensity images near 6173 \AA\ from SDO. This investigation aims to improve our understanding of energy release and transportation in the lower atmosphere of solar flares. The results can also provide insights into the flares on solar-type stars, which are often detected in WL, whereas their \lya\ emission is difficult to be observed because of interstellar absorption and scattering \citep{Linsky2014}. We introduce the observational data and methods in Section \ref{sec:data_methods} and present the analysis and results in Section \ref{sec:results}. A summary and discussion are provided in Sections \ref{sec:summary} and \ref{sec:discussion}, respectively.

\section{Observational Data and Methods}
\label{sec:data_methods}

\subsection{Instruments and Data}

Three spaceborne instruments are used for this statistical study: the X-Ray Sensor \citep[XRS;][]{Hanser1996} and the EUVS on board the GOES-N series, as well as the Helioseismic and Magnetic Imager \citep[HMI;][]{Scherrer2012} on board SDO. XRS measures the solar irradiance in two SXR wave bands, 1--8 \AA\ and 0.5--4 \AA, with a time cadence of $\sim$2 s. The GOES flare class, from A to X, is defined based on the peak flux of the 1--8 \AA\ channel. The EUVS measures the solar extreme ultraviolet irradiance in five channels (A--E) spanning approximately 50 to 1270 \AA. The \lya\ data are obtained from channel E, which covers a $\sim$90 \AA\ (1180--1270 \AA) bandpass around the \lya\ line at 1216 \AA, with a cadence of $\sim$10 s. In addition, we obtain the pseudo-continuum intensity images observed near the photospheric \ion{Fe}{1} 6173 \AA\ line from HMI. These imaging observations typically feature a pixel size of $\sim$0.5\arcsec\ and a cadence of 45 s. In this study, we specifically employ pre-processed level 1.5 HMI data, which have a pixel scale of $\sim$0.6\arcsec. The HMI data were corrected for solar differential rotation using the \texttt{drotmap.pro} routine available in SSWIDL \citep{SSW1998}.

\subsection{WLF Data Set}
\label{sec:wlf_dataset}  

The WLF data set used here was compiled from some previous statistical surveys \citep{Buitrago-Casas2015,Kuhar2016,Huang2016,Namekata2017,Watanabe2017,Song2018jj,Song2018aa,Castellanos2020} and case studies \citep{Hao2017,Song2018ApJb,Watanabe2020,Lee2017}. Some of these WLF studies used not only HMI data but also WL continuum measurements from the Michelson Doppler Imager \citep[MDI;][]{Scherrer1995} on board the Solar and Heliospheric Observatory \citep[SOHO;][]{SOHO1995}. As WL continuum passbands differ among instruments, here we restrict the sample to WLFs observed by HMI.

\begin{figure}[h]
	\centering
	\includegraphics[width=0.55\textwidth]{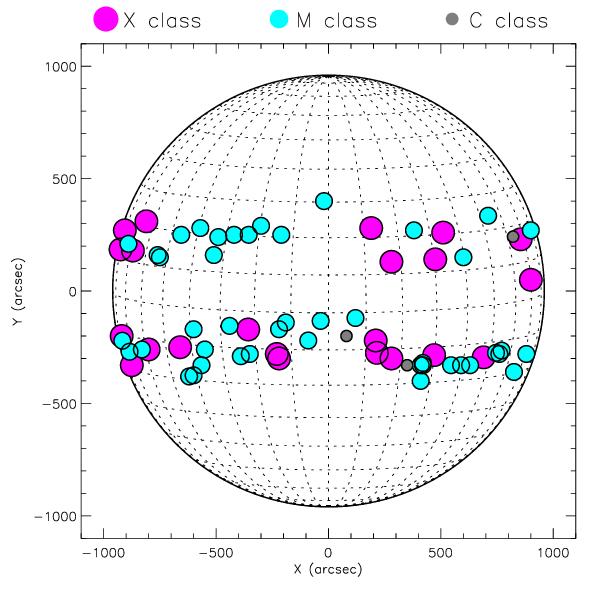}
	\caption{Spatial distribution of the 69 WLFs. Magenta, cyan, and gray circles represent X-, M-, and C-class flares, respectively.}
	\label{wllya_fig1}
\end{figure}

Events were excluded if the WL enhancement had low signal-to-noise ratio, if the \lya\ data were contaminated by geocoronal absorption \citep{Baliukin2019}, or if the light curves were affected by overlapping flares. The final data set comprises 69 WLFs that occurred between 2010 October 16 and 2015 October 2, corresponding to the ascending and maximum phases of solar cycle 24. The sample includes 3 C-class, 44 M-class, and 22 X-class events, with GOES classifications from C7.7 to X6.9 (see Table \ref{flarelist}). Figure \ref{wllya_fig1} shows the spatial distribution of the 69 WLFs on solar disk. In latitude, most events (42; $\sim$61\%) occurred in the southern hemisphere, consistent with the north–south asymmetry reported for GOES SXR flares during solar cycle 24 \citep{Joshi2019}. In longitude, roughly half (38; $\sim$55\%) were located in the eastern hemisphere, indicating no strong longitudinal preference overall.

\subsection{Data Reduction and Parameter Measurement}
\label{paracal}

\begin{figure}[htb]
	\centering
	\includegraphics[width=0.8\textwidth]{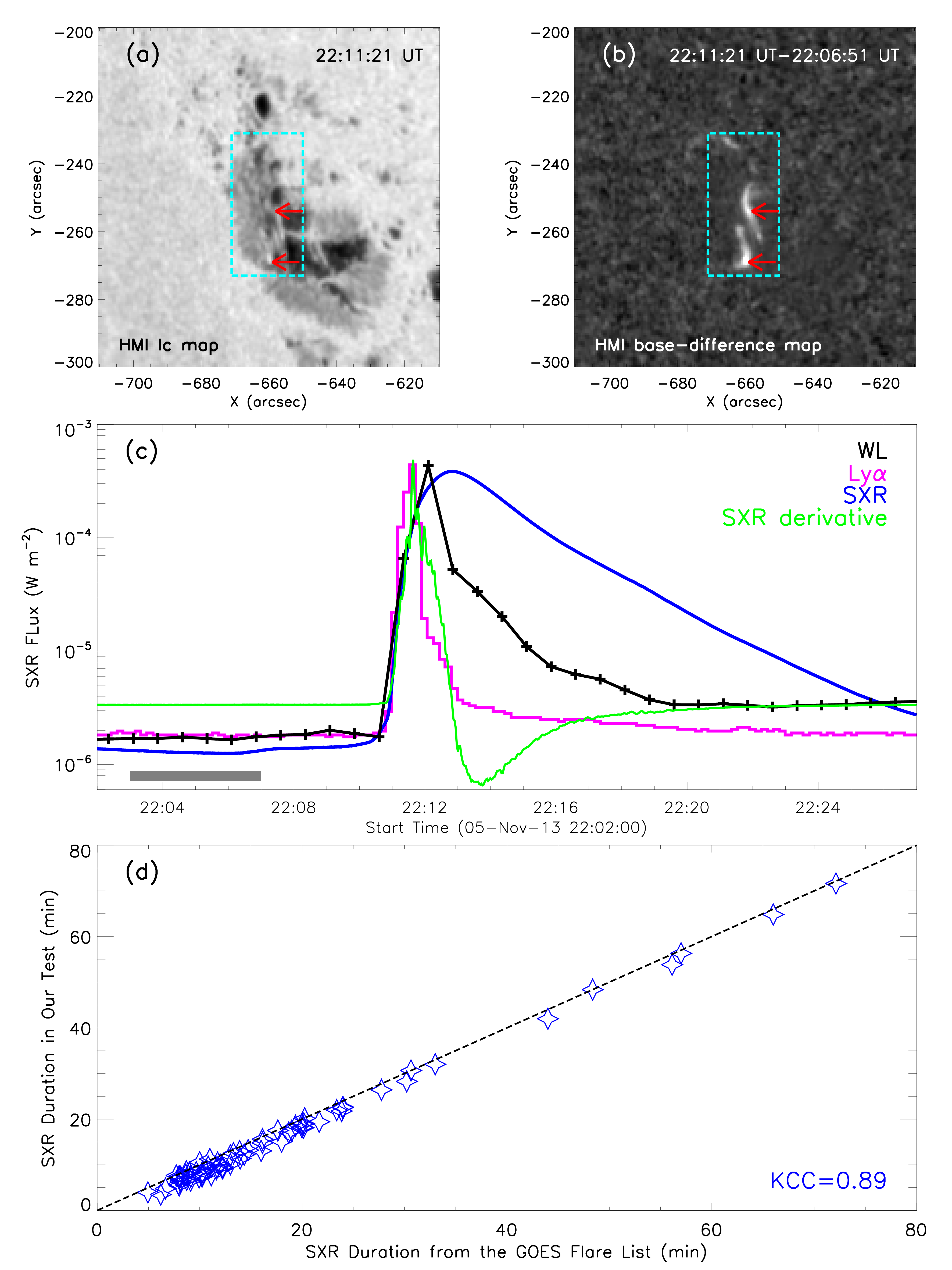}
	\caption{Example of a WLF (\# 42 in Table \ref{flarelist}) and a comparison of the SXR duration based on different pre-flare backgrounds ($F_{\text{bkg}}$). (a) and (b) The original HMI continuum image and the base-difference image, respectively, of the X3.3 flare on 2013 November 5. The cyan box indicates the spatial integration region for the WL emission curve (black line in panel (c)). Red arrows point to the two bright flare ribbons in WL images. (c) Light curves of WL (black), \lya\ (magenta), SXR (blue), and SXR time derivative (green). Note that the time profiles of WL, \lya, and the SXR time derivative are plotted in an arbitrary scale. The four-minute $F_{\text{bkg}}$ interval used in this study is marked by the gray bar in the bottom-left corner. (d) Comparison of the SXR duration estimated using our pre-flare background selection (y-axis) versus the officially recorded GOES duration (x-axis). Their linear Kendall's Tau correlation coefficient (KCC) is noted in the bottom-right corner. The black dashed line is the 1:1 reference line. Note that both SXR durations are computed using a half-maximum threshold, which is solely to validate the background selection.}
	\label{wllya_fig2}
\end{figure}

The light curves of solar flares at different wavelengths reflect distinct physical processes. Disk-integrated (irradiance) light curves in \lya\ and SXR 1--8 \AA\ (hereafter SXR) bands are taken directly from GOES. For the WL band, we follow \citet{Hudson2006} and construct the flare light curve (see the black line in Figure~\ref{wllya_fig2}(c)) by integrating the HMI continuum intensity over the flare region (indicated by the cyan dashed box in Figures~\ref{wllya_fig2}(a) and (b)). Following \citet{Namekata2017}, the WL curve is linearly interpolated to a 1 s cadence to improve the precision of the measurements.

\subsubsection{Data Preprocessing}
\label{Radiometric}

The GOES data in the \lya\ and SXR bands are radiometrically calibrated, with a physical unit of $\mathrm{W\,m^{-2}}$. 
We then convert the GOES flux ($F_{\text{ori}}(t)$) to an equivalent hemispheric radiative power (\citealt{Milligan2020,Greatorex2024}):
\begin{equation}
F_{\text{GOES}} = F_{\text{ori}}\,\times\,2\pi d^{2}\,\times\,10^{7}  \quad (\mathrm{erg~s^{-1}}),
\end{equation}
where $d$ is the Sun-Earth distance (i.e., 1 au) and $10^7$ converts $\mathrm{W}$ to $\mathrm{erg\,s^{-1}}$.

HMI does not provide radiometrically calibrated data. Here we calibrate the HMI data using the method described in \citet{Kleint2016} (also see \citealt{Heinzel2017,jejcic18,Castellanos2020,GarciaRivas2024}). Specifically, we measure the average count in a quiet-Sun region near disk center ($\approx 60000~\mathrm{DN}$) and compare it to the solar atlas continuum at 6173~\AA\ \citep{Neckel1994}, $0.315 \times 10^{7}\ \mathrm{erg\,s^{-1}\,cm^{-2}\,sr^{-1}\,\AA^{-1}}$, which gives a conversion factor $C_{\mathrm{HMI}}$ $\approx 52.5\ \mathrm{erg\,s^{-1}\,cm^{-2}\,sr^{-1}\,\AA^{-1}\,DN^{-1}}$. 
By applying this factor, we convert the integrated HMI flux over the flare region,
$F_{\text{DN}}(t)=\sum_i \mathrm{DN}_i(t)$ ($i$ represents the pixel within the flare region), into physical units:
\begin{equation}
F_{\text{HMI}} = F_{\text{DN}} \cdot C_{\mathrm{HMI}} \quad (\mathrm{erg\,s^{-1}\,cm^{-2}\,sr^{-1}\,\AA^{-1}}).
\end{equation}

We assume that the HMI continuum emission is isotropic per unit solid angle and consider radiation emitted only into the outward hemisphere. Under these assumptions, the hemispheric radiative power ($F_{\text{HMI}}$) of a flare region across the bandpass $\Delta\lambda$ can be expressed as:
\begin{equation}
F_{\text{HMI}} = F_{\text{DN}} \cdot C_{\mathrm{HMI}} \cdot A_{\mathrm{pix}} \cdot \pi  \cdot \Delta\lambda \quad (\mathrm{erg\,s^{-1}}).
\end{equation}
Here, $A_{\mathrm{pix}}$ denotes the area of a single HMI pixel, calculated as $(0.6 \times 725 \times 10^5)^2~\mathrm{cm^2}$. We further assume that the HMI WL continuum intensity represents solely the average intensity of the WL continuum near 6173 \AA. Therefore, we conservatively set $\Delta\lambda$ to 1 \AA\ in the following parameter measurements.

\subsubsection{Parameter Definition and Measurement}

Using the calibrated SXR, \lya, and WL fluxes, we measure the following parameters.

\begin{itemize} 
	
	\item \textbf{Flux Enhancement and \lya\ Contrast}
    
The flux enhancement ($F_{\text{enhancement}}$) or emission increase in each band is quantified by the peak flux ($F_{\text{peak}}$) and the pre-flare background flux ($F_{\text{bkg}}$), namely
\begin{equation}
F_{\text{enhancement}} = F_{\text{peak}} - F_{\text{bkg}} \quad (\mathrm{erg~s^{-1}}),
\end{equation}
where $F_{\text{bkg}}$ is defined as the mean flux over the 4-minute interval (indicated by the gray bar in Figure~\ref{wllya_fig2}(c)) preceding the GOES start time for each band. Here $F_{\text{enhancement}}$ represents the absolute increase in radiated power.

For the \lya\ band, we also calculate the relative enhancement percentage (i.e., contrast) of the flare emission, defined as $(F_{\text{enhancement}}/F_{\text{bkg}}) \times 100\%$, following previous studies \citep[e.g.,][]{Milligan2020,Milligan2021}.

	\item \textbf{Peak Time and Evolution Timescales}
    
Temporal parameters include the peak time ($t_{\text{peak}}$), rise time ($t_{\text{rise}}$), decay time ($t_{\text{decay}}$), and duration ($\tau$). $t_{\text{peak}}$ in a given band is the time when the flux reaches its maximum. $t_{\text{rise}}$ is measured from a start time to $t_{\text{peak}}$. The start time is defined as the time when the flux starts to exceed $F_{\text{bkg}}$ by $3\sigma$. $t_{\text{decay}}$ is measured from $t_{\text{peak}}$ to the end time when the flux decays to $1/e$ of $F_{\text{enhancement}}$. The $e$-folding threshold mitigates an underestimation of $t_{\text{decay}}$ compared to a half-maximum threshold and facilitates direct comparisons with stellar-flare studies \citep[e.g.,][]{Maehara2015,Namekata2017,Kowalski2024}. $\tau$ is the period from the start to end times, i.e., $\tau = t_{\text{rise}} + t_{\text{decay}}$.
	
Note that the duration $\tau$ defined here is somewhat different from some other definitions. For example, for the SXR duration, according to the GOES criteria\footnote{\url{https://www.swpc.noaa.gov/products/goes-x-ray-flux}}, the flare start is defined as the first of four consecutive minutes of monotonically increasing SXR 1--8 \AA\ flux, and the end is defined as the time when the flux decays to halfway between the maximum and the pre-flare background. Here we compared the SXR durations derived from this half-maximum criterion using our four-minute $F_{\text{bkg}}$ with the official GOES durations (Figure \ref{wllya_fig2}(d)). As shown, the two agree closely with a Kendall's Tau correlation coefficient (KCC) of 0.89, i.e., a very high correlation\footnote{According to \citet{kuckartz2013statistik}, 
$|\mathrm{KCC}|<0.1$ indicates no correlation, $0.1\le|\mathrm{KCC}|<0.3$ denotes a low correlation, $0.3\le|\mathrm{KCC}|<0.5$ signifies a medium correlation, $0.5\le|\mathrm{KCC}|<0.7$ represents a high correlation, and $0.7\le|\mathrm{KCC}|\le1$ corresponds to a very high correlation.}, supporting the reliability of our method. In this paper, unless otherwise stated, all reported correlation coefficients are evaluated in the log10--log10 scale and are quantified using the nonparametric Kendall's Tau correlation method, which does not rely on any assumptions of the data distribution.

	\item \textbf{Radiated Energy}\\
The radiated energy ($E$) is the time integral of the background-subtracted radiated power over the flare duration, 
        \begin{equation}
    E = \int_{t_{\text{start}}}^{t_{\text{end}}} (F(t) - F_{\text{bkg}}) \, dt \quad (\mathrm{erg}).
    \end{equation}
In this equation, $F(t)$ represents the radiometrically calibrated fluxes, i.e., $F_{\text{GOES}}$ or $F_{\text{HMI}}$ (see Section \ref{Radiometric}). It should be noted that the integration limits, $t_{\text{start}}$ and $t_{\text{end}}$ ($t_{\text{end}} - t_{\text{start}} = \tau$, namely the duration as described above), are determined independently for each waveband. Using this method, we obtain $E_{\text{\lya}}$, $E_{\text{WL}}$, and $E_{\text{SXR}}$ (see Table \ref{flarelist}).
For $E_{\text{WL}}$, it is crucial to emphasize that this value represents the energy radiated specifically in the narrow-band continuum with a width of 1 \AA\ (i.e., $\Delta\lambda=1\,\mathrm{\AA}$) at 6173 \AA.
As a rough estimate, the total radiated energy in the broadband WL continuum (3800--7600 \AA) can be obtained by multiplying the $E_{\text{WL}}$ listed in Table \ref{flarelist} by the continuum width (3800 \AA), although this assumes a flat spectrum.

\end{itemize}

\begin{longtable}{lccccccc} 
\caption{Information of the 69 WLFs \label{flarelist}}\\ 

\hline
\hline 
\# & Date & SXR   & GOES & Location & $E_{\text{Ly}_{\alpha}}$ & $E_{\text{WL}}$ & $E_{\text{SXR 1--8 \AA}}$ \\
 &  & Peak Time & Class &  &   &  &  \\
 & (yyyy-mm-dd) & (UT) &  & (arcsec) &  ($10^{29}$ erg) & ($10^{26}$ erg) & ($10^{29}$ erg) \\
		\midrule
\endfirsthead 

\multicolumn{8}{c}
{{\tablename\ \thetable{} Information of the 69 WLFs}} \\ 
\hline\hline 
\# & Date & SXR   & GOES & Location & \bf $E_{\text{Ly}_{\alpha}}$ & $E_{\text{WL}}$ & $E_{\text{SXR 1--8 \AA}}$ \\
 &  & Peak Time & Class &  &   &  &  \\
 & (yyyy-mm-dd) & (UT) &  & (arcsec) &  ($10^{29}$ erg) & ($10^{26}$ erg) & ($10^{29}$ erg) \\
		\midrule
\endhead

\hline\hline 
\endlastfoot 
1     & 2010-10-16 & 19:12 & M2.9  & (410, $-$400) & 0.32 & 0.70 & 0.09 \\
2     & 2011-02-13 & 17:38 & M6.6  & ($-$90, $-$220) &  1.28 & 2.65 & 0.66 \\
3     & 2011-02-15 & 01:56 & X2.2  & (210, $-$220) &  7.16 & 4.14 & 2.32 \\
4     & 2011-02-18 & 10:11 & M6.6  & (745, $-$275) &  0.45 & 0.75 & 0.23 \\
5     & 2011-02-18 & 13:03 & M1.4  & (760, $-$280) &  0.17 & 0.29 & 0.04 \\
6     & 2011-03-09 & 23:23 & X1.5  & (190, 280) &  1.69 & 0.34 & 0.92 \\
7     & 2011-03-14 & 19:52 & M4.2  & (710, 335) &  0.71 & 0.40 & 0.13 \\
8     & 2011-07-30 & 02:09 & M9.3  & ($-$510, 160) &  1.28 & 1.54 & 0.27 \\
9     & 2011-08-08 & 22:09 & C7.7  & (821, 242) &  0.05 & 0.55 & 0.09 \\
10    & 2011-08-09 & 08:05 & X6.9  & (855, 230) &  2.07 & 4.28 & 2.48 \\
11    & 2011-09-06 & 22:20 & X2.1  & (280, 130) &  3.35 & 2.05 & 0.85 \\
12    & 2011-09-07 & 22:38 & X1.8  & (475, 141) &  4.89 & 3.11 & 0.95 \\
13    & 2011-09-08 & 15:46 & M6.7  & (600, 150) &  7.29 & 2.43 & 0.59 \\
14    & 2011-09-24 & 20:36 & M5.8  & ($-$750, 150) &  0.42 & 0.23 & 0.43 \\
15    & 2011-09-24 & 17:25 & M3.1  & ($-$760, 160) &  0.04 & 0.26 & 0.13 \\
16    & 2011-11-03 & 20:27 & X1.9  & ($-$810, 310) &  2.94 & 5.33 & 1.43 \\
17    & 2011-12-26 & 20:30 & M2.3  & (630, $-$330) &  0.16 & 0.66 & 0.29 \\
18    & 2011-12-31 & 13:15 & M2.4  & ($-$620, $-$380) &  0.47 & 0.45 & 0.10 \\
19    & 2011-12-31 & 16:26 & M1.5  & ($-$600, $-$375) &  0.81 & 0.68 & 0.12 \\
20    & 2012-03-09 & 03:53 & M6.3  & ($-$20, 400) &  4.85 & 5.84 & 2.02 \\
21    & 2012-05-06 & 01:18 & M1.1  & ($-$890, 210) &  0.27 & 0.78 & 0.03 \\
22    & 2012-05-08 & 13:08 & M1.4  & ($-$655, 250) &  1.41 & 0.42 & 0.06 \\
23    & 2012-05-09 & 12:32 & M4.7  & ($-$490, 240) &  5.66 & 2.30 & 0.24 \\
24    & 2012-05-09 & 21:05 & M4.1  & ($-$420, 250) &  0.46 & 1.63 & 0.18 \\
25    & 2012-05-10 & 04:18 & M5.7  & ($-$355, 250) &  9.09 & 1.29 & 0.29 \\
26    & 2012-05-10 & 20:26 & M1.7  & ($-$210, 250) &  2.75 & 0.79 & 0.08 \\
27    & 2012-06-03 & 17:55 & M3.3  & ($-$570, 280) &  1.31 & 0.56 & 0.09 \\
28    & 2012-06-09 & 16:53 & M1.8  & ($-$885, $-$270) &  0.54 & 0.44 & 0.09 \\
29    & 2012-07-04 & 21:27 & C9.5  & (350, $-$330) &  0.25 & 0.19 & 0.01 \\
30    & 2012-07-04 & 22:09 & M4.6  & (420, $-$320) &  1.39 & 0.43 & 0.25 \\
31    & 2012-07-05 & 02:42 & M2.2  & (420, $-$330) &  0.73 & 1.85 & 0.09 \\
32    & 2012-07-05 & 03:36 & M4.7  & (410, $-$330) &  0.89 & 0.77 & 0.13 \\
33    & 2012-07-05 & 20:14 & M1.6  & (545, $-$330) &  1.87 & 0.27 & 0.15 \\
34    & 2012-07-06 & 01:40 & M2.9  & (590, $-$330) &  0.09 & 0.50 & 0.06 \\
35    & 2012-08-06 & 04:38 & M1.6  & ($-$917, $-$221) &  0.09 & 0.11 & 0.05 \\
36    & 2012-10-22 & 18:51 & M5.0  & ($-$830, $-$260) &  2.32 & 1.24 & 0.53 \\
37    & 2012-10-23 & 03:17 & X1.8  & ($-$800, $-$260) &  0.87 & 1.69 & 0.67 \\
38    & 2013-05-13 & 16:05 & X2.8  & ($-$925, 185) &  0.32 & 4.51 & 3.47 \\
39    & 2013-05-15 & 01:48 & X1.2  & ($-$870, 180) &  2.93 & 3.72 & 1.75 \\
40    & 2013-07-08 & 01:22 & C9.7  & (80, $-$200) &  1.20 & 0.27 & 0.02 \\
41    & 2013-10-28 & 02:03 & X1.0  & (900, 50) &  3.47 & 0.48 & 1.17 \\
42    & 2013-11-05 & 22:12 & X3.3  & ($-$660, $-$250) &  1.20 & 3.66 & 0.82 \\
43    & 2013-11-06 & 13:46 & M3.8  & ($-$550, $-$260) &  2.50 & 2.24 & 0.24 \\
44    & 2013-11-07 & 14:25 & M2.4  & ($-$350, $-$280) &  1.66 & 0.51 & 0.15 \\
45    & 2013-11-08 & 04:26 & X1.1  & ($-$230, $-$280) &  1.40 & 1.76 & 0.36 \\
46    & 2013-11-10 & 05:14 & X1.1  & (215, $-$275) &  3.16 & 1.14 & 0.47 \\
47    & 2013-12-22 & 15:12 & M3.3  & (770, $-$265) &  0.25 & 1.11 & 0.17 \\
48    & 2014-01-07 & 10:13 & M7.2  & ($-$220, $-$170) &  4.64 & 2.00 & 1.31 \\
49    & 2014-02-02 & 06:34 & M2.6  & ($-$300, 290) &  1.01 & 0.97 & 0.10 \\
50    & 2014-02-04 & 04:00 & M5.2  & (120, $-$120) &  0.57 & 0.58 & 0.23 \\
51    & 2014-02-25 & 00:49 & X4.9  & ($-$920, $-$200) &  4.93 & 5.52 & 6.25 \\
52    & 2014-03-12 & 22:34 & M9.3  & (900, 270) &  0.27 & 0.57 & 0.45 \\
53    & 2014-03-28 & 23:51 & M2.6  & (380, 270) &  2.52 & 0.67 & 0.21 \\
54    & 2014-03-29 & 17:48 & X1.0  & (510, 260) &  2.49 & 1.10 & 0.59 \\
55    & 2014-06-10 & 11:42 & X2.2  & ($-$875, $-$330) &  0.70 & 0.94 & 0.62 \\
56    & 2014-10-20 & 19:02 & M1.4  & ($-$565, $-$330) &  0.75 & 0.21 & 0.05 \\
57    & 2014-10-22 & 01:59 & M8.7  & ($-$390, $-$290) &  1.61 & 0.18 & 2.94 \\
58    & 2014-10-22 & 14:28 & X1.6  & ($-$220, $-$300) &  1.47 & 0.94 & 4.80 \\
59    & 2014-10-24 & 21:41 & X3.1  & (280, $-$300) &  36.48 & 5.08 & 12.57 \\
60    & 2014-10-27 & 14:47 & X2.0  & (690, $-$295) &  8.08 & 0.75 & 6.51 \\
61    & 2014-12-20 & 00:28 & X1.8  & (470, $-$285) &  27.37 & 1.76 & 5.07 \\
62    & 2015-03-10 & 03:24 & M5.1  & ($-$600, $-$170) &  0.99 & 0.61 & 0.25 \\
63    & 2015-03-10 & 00:02 & M2.9  & ($-$440, $-$155) &  0.40 & 0.10 & 0.13 \\
64    & 2015-03-11 & 16:22 & X2.1  & ($-$357, $-$171) &  5.97 & 2.70 & 1.77 \\
65    & 2015-03-12 & 04:46 & M3.2  & ($-$190, $-$140) &  0.78 & 0.38 & 0.16 \\
66    & 2015-03-12 & 21:51 & M2.7  & ($-$35, $-$133) &  0.51 & 0.10 & 0.15 \\
67    & 2015-05-05 & 22:11 & X2.7  & ($-$906, 270) &  2.34 & 3.45 & 1.22 \\
68    & 2015-08-28 & 19:03 & M2.1  & (880, $-$280) &  0.50 & 0.15 & 0.06 \\
69    & 2015-10-02 & 00:13 & M5.5  & (825, $-$360) &  4.52 & 0.87 & 0.26 \\
\end{longtable}
\begin{minipage}{\linewidth}
\footnotesize
\textit{Note.} $E_{\text{\lya}}$ and $E_{\text{SXR}}$ are derived from GOES observations, corresponding to the EUVS-E bandpass (1180--1270~\AA) around \lya\ and the XRS 1--8~\AA\ band, respectively. $E_{\text{WL}}$ is estimated from the HMI continuum data near 6173~\AA, assuming $\Delta\lambda=1$~\AA.
\end{minipage}

\section{Analysis and Results}
\label{sec:results}

\subsection{\lya\ Contrast in WLFs}

\begin{figure}[htb]
	\centering
	\includegraphics[width=0.6\textwidth]{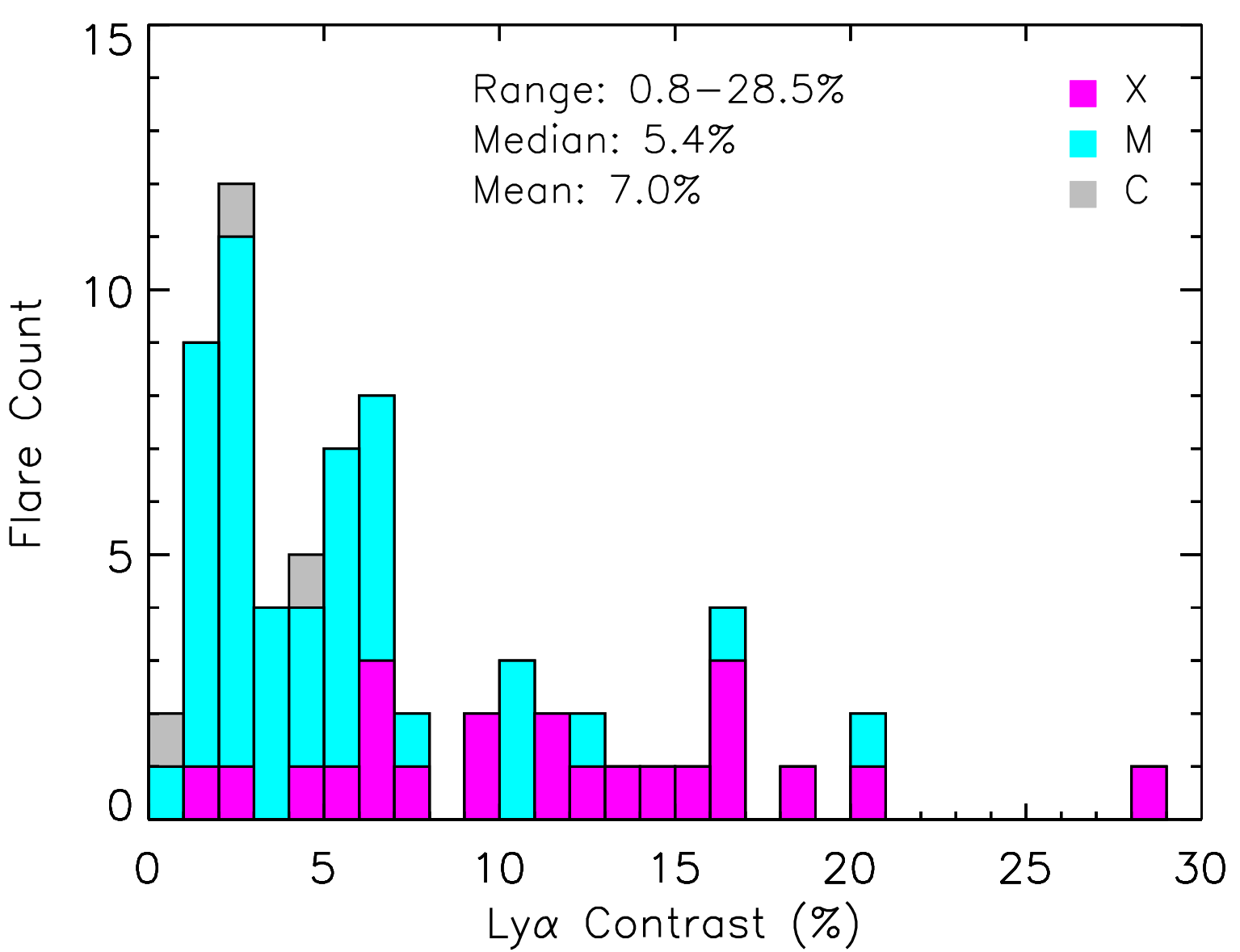}
	\caption{Distribution of the \lya\ contrast. Magenta, cyan, and gray bars represent X-, M-, and C-class flares, respectively. The range, median, and mean of the \lya\ contrast are annotated in the figure.}
	\label{wllya_fig3}
\end{figure}

We first calculated the \lya\ contrast for the 69 WLFs. The distribution histogram is shown in Figure \ref{wllya_fig3}. Quantitatively, the \lya\ contrast of these WLFs ranges from 0.8\% to 28.5\%, with a median of 5.4\% and a mean of 7.0\%. The distribution shows that high \lya\ contrast values are predominantly associated with X- and M-class flares. However, some C-class flares exhibit a contrast greater than that of some M-class or even X-class flares.
In addition, we performed a two-sample Anderson-Darling test to explore the difference in the \lya\ contrast between M- and X-class flares. The test yields a statistic of 11.70 and a p-value of $<$0.001. This suggests that the \lya\ contrast values of X-class flares are also significantly distinct from those of M-class flares.

\subsection{Statistical Relationships of Flux Enhancements}

\begin{figure}[htb]
	\centering
	\includegraphics[width=1.0\textwidth]{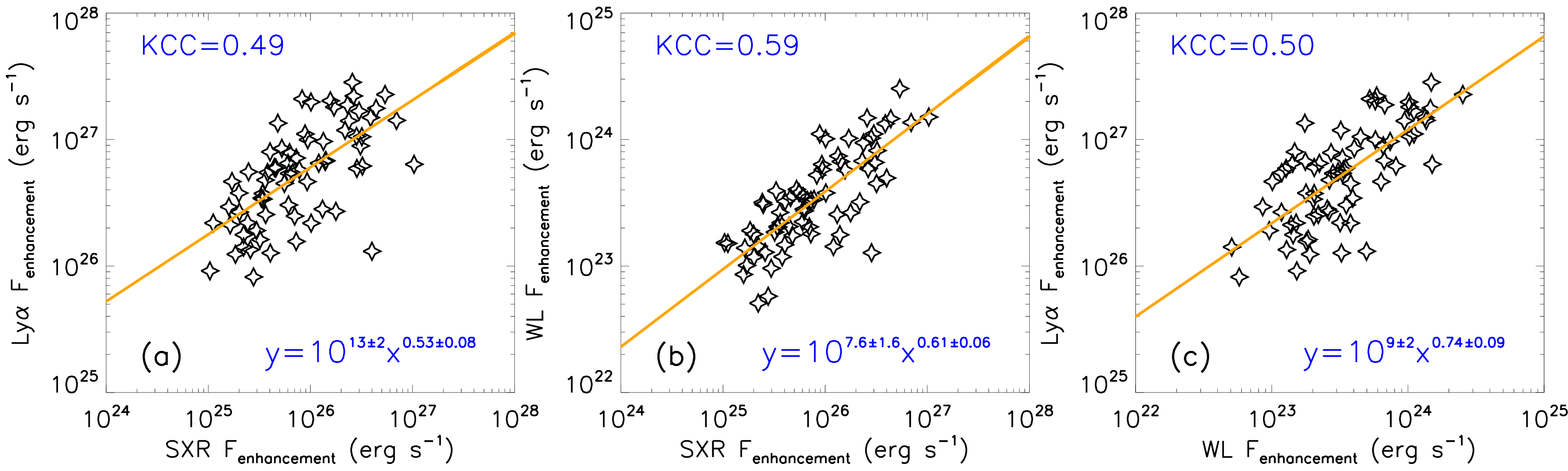}
	\caption{Relationships between peak enhancements ($F_{\text{enhancement}}$) in \lya, WL, and SXR bands. Here the WL enhancement represents the emission increase in the HMI continuum near 6173~\AA\ with an assumed bandpass of $\Delta\lambda=1$~\AA. The gray dashed line and the orange solid line in each panel represent the 1:1 reference and the power-law best fit, respectively. The corresponding Kendall's Tau correlation coefficient (KCC) and power-law fitting result are annotated in the top-left and bottom-right corners, respectively.}
	\label{wllya_fig4}
\end{figure}

The background-subtracted peak enhancement, $F_{enhancement}$, is a useful proxy for the radiative output in a given band. Figure~\ref{wllya_fig4} shows the scatter plots among the \lya, WL, and SXR peak enhancements for the 69 WLFs. From Figures~\ref{wllya_fig4}(a) and (b), we can see that the \lya\ and WL enhancements exhibit medium to high positive power-law correlations with the SXR enhancement, with KCCs of $0.49$ and $0.59$, respectively. A clear positive power-law relation is also found between the \lya\ and WL enhancements (KCC $= 0.50$, Figure~\ref{wllya_fig4}(c)). These results indicate that the \lya\ and WL emissions respond strongly to flare heating and scale with the overall flare magnitude represented by SXR enhancement, particularly for the WL band.

\subsection{Statistical Relationships of Peak Time Differences}
 
\begin{figure}[htb]
	\centering
	\includegraphics[width=1.0\textwidth]{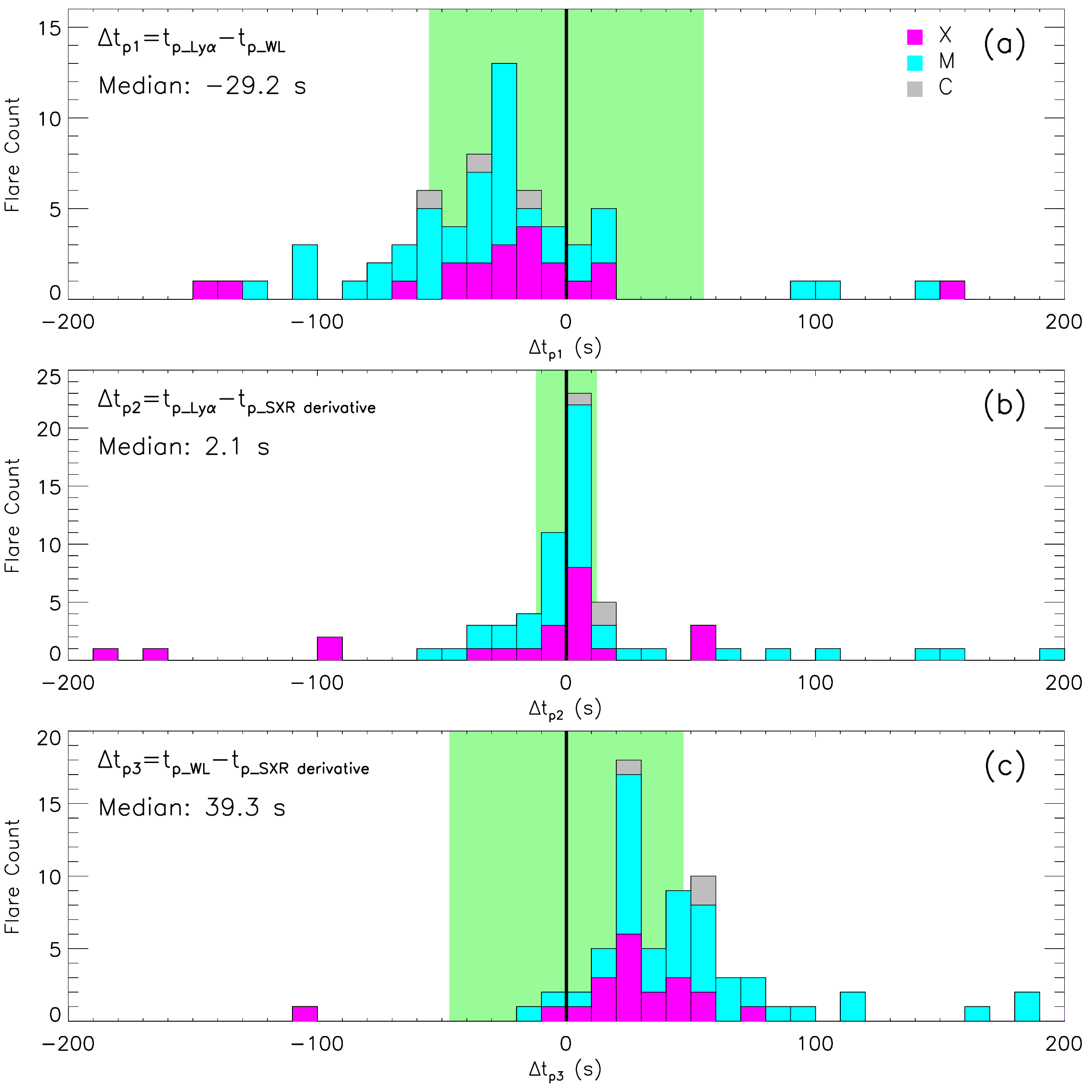}
	\caption{Distributions of the peak time ($t_{\text{p}}$) differences among \lya, WL, and the SXR time derivative. Magenta, cyan, and gray bars represent X-, M-, and C-class flares, respectively. To clearly show the distribution, the x-axis range in all panels is limited to $\pm$200 s, which covers the vast majority of the flares ($>$94\%), and the few events outside this range do not affect the overall statistical results. The black vertical line in each panel marks $\Delta t_{\text{p}} = 0$, and the green shaded region represents the maximum temporal uncertainty. The uncertainties for $t_{\text{p1}}$, $t_{\text{p2}}$, and $t_{\text{p3}}$ are 55 s, 12 s, and 47 s, respectively.}
	\label{wllya_fig5}
\end{figure}

Based on the light curves, we calculated the time differences between the peak times of \lya, WL, and the SXR time derivative (hereafter the SXR derivative). The distributions are shown in Figure \ref{wllya_fig5}, where $\Delta t_{\text{p1}}=t_{\text{p}\_\text{Ly}\alpha}-t_{\text{p}\_\text{WL}}$, $\Delta t_{\text{p2}}=t_{\text{p}\_\text{Ly}\alpha}-t_{\text{p}\_\text{SXR derivative}}$, and $\Delta t_{\text{p3}}=t_{\text{p}\_\text{WL}}-t_{\text{p}\_\text{SXR derivative}}$ represent the time lags between \lya\ and WL, \lya\ and SXR derivative, and WL and SXR derivative, respectively. Based on the observational time resolutions of the three bands, the maximum uncertainties for $\Delta t_{\text{p1}}$, $\Delta t_{\text{p2}}$, and $\Delta t_{\text{p3}}$ are approximately 55 s, 12 s, and 47 s, respectively, as indicated by the green shaded regions in Figure \ref{wllya_fig5}.

Figure \ref{wllya_fig5}(a) shows that the distribution of $\Delta t_{\text{p1}}$ is skewed toward negative values, with a median of $-$29.2 s. For 56 flares ($\sim$81\%), $\Delta t_{\text{p1}} \le 0$ s, and for 46 WLFs ($\sim$67\%), $\Delta t_{\text{p1}}$ falls within the temporal uncertainty range. This indicates that in most WLFs, the \lya\ emission peaks at the same time as or earlier than the WL emission. Figure \ref{wllya_fig5}(b) presents the time lag between \lya\ and SXR derivative ($\Delta t_{\text{p2}}$). The distribution is concentrated near 0 s (median = 2.1 s), with 36 flares (52\%) falling within the uncertainty range, i.e., holding the Neupert effect \citep{Neupert1968}. This timing relationship highlights the nonthermal origin of flare \lya\ emission. The relationship between the WL and SXR derivative peak times ($\Delta t_{\text{p3}}$, Figure \ref{wllya_fig5}(c)) is consistent with the findings from Figures \ref{wllya_fig5}(a) and (b), indicating that the WL emission peaks at the same time as or later than the SXR derivative. Specifically, 41 flares (59\%) peak within the uncertainty range, while for 26 flares (38\%), the WL peak lags behind the SXR derivative peak.

\subsection{Statistical Relationships of Evolution Timescales}
 
\begin{figure}[htb]
	\centering
	\includegraphics[width=0.6\textwidth]{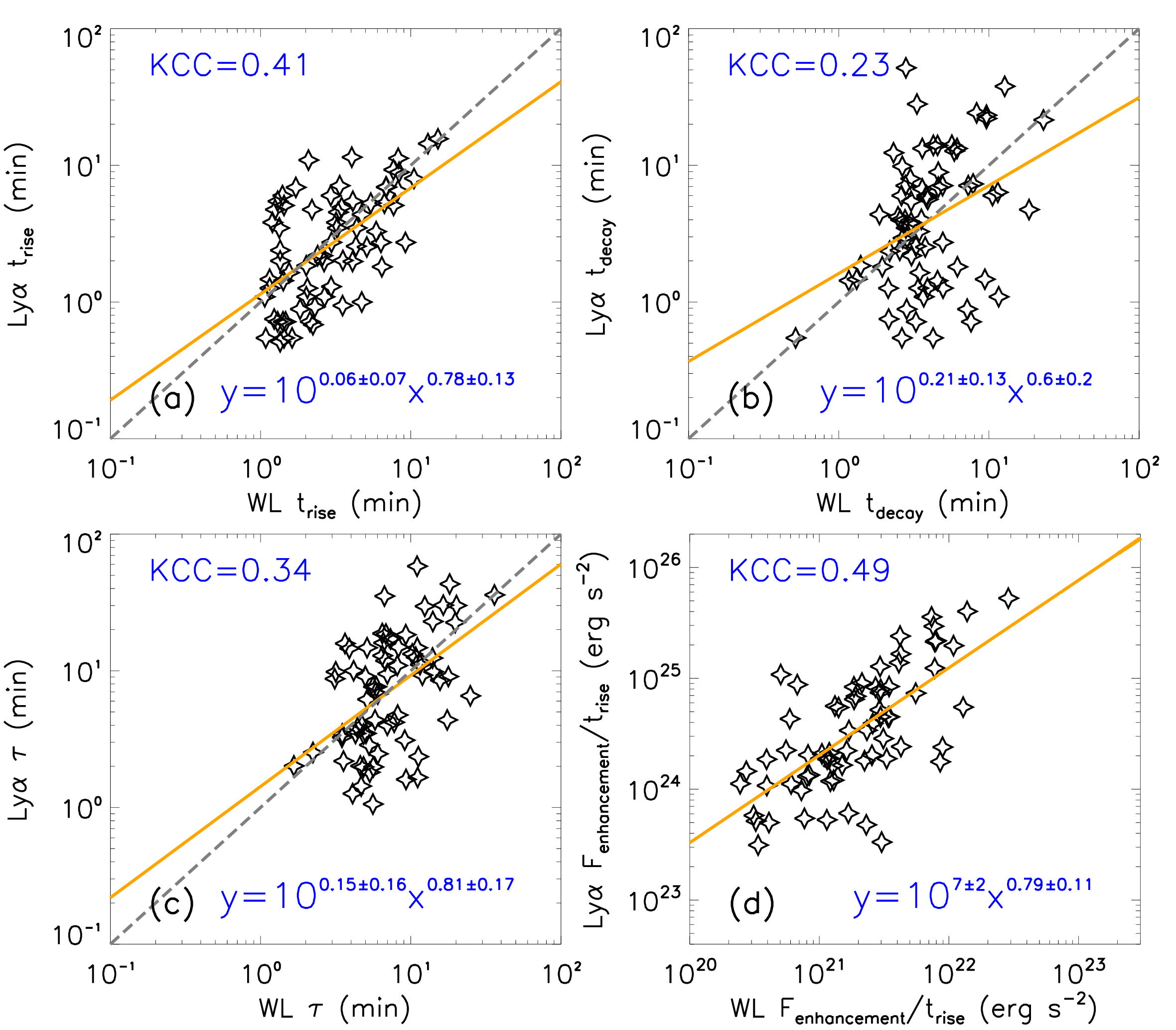}
	\caption{Relationships between \lya\ and WL for (a) rise time ($t_{\text{rise}}$), (b) decay time ($t_{\text{decay}}$), (c) duration ($\tau$), and (d) increase rate of peak enhancement ($F_{\text{enhancement}}/t_{\text{rise}}$). Here the WL enhancement denotes the emission increase in the HMI continuum near 6173~\AA\ with an assumed bandpass of $\Delta\lambda=1$~\AA. The gray dashed line and the orange solid line in each panel represent the 1:1 reference and the power-law best fit, respectively. The corresponding KCC and power-law fitting result are annotated in the top-left and bottom-right corners, respectively.}
	\label{wllya_fig6}
\end{figure}

We also examined the relationships between the \lya\ and WL emissions in $t_{\text{rise}}$, $t_{\text{decay}}$, $\tau$, and the increase rate of peak enhancement, i.e., $F_{\text{enhancement}}/t_{\text{rise}}$, as shown in Figure~\ref{wllya_fig6}. The increase rate of peak enhancement, representing the slope of the rise phase, reflects the impulsiveness of the flare emission. From Figures \ref{wllya_fig6}(a)--(c), we find that the \lya\ and WL emissions are positively correlated in $t_{\text{rise}}$ (KCC $= 0.41$) but show low or medium correlations in $t_{\text{decay}}$ and $\tau$ (KCCs $\le 0.34$). We further find that the increase rate of peak enhancement in \lya\ and WL exhibits a better positive power-law correlation (KCC $= 0.49$; Figure \ref{wllya_fig6}(d)), and that the increase rate in WL is generally greater than that in \lya.

Quantitatively, the rise times of \lya\ and WL are highly consistent, both with median and mean values of $\sim$3 minutes and $\sim$4 minutes, respectively (see Table~\ref{wllyatab2}). In addition, we compare the \lya\ and WL rise times from this study with previous statistical results for solar and stellar flares \citep{Lu2021catalog,Yan2021MNRAS} in Table~\ref{wllyatab2}. It can be seen that the \lya\ $t_{\text{rise}}$ in WLFs is shorter (i.e., more impulsive) than that found in the general flare catalog that includes both WLFs and non-WLFs. For WL emission, the rise time of stellar flares on solar-type stars is longer than that of solar WLFs, which suggests a longer-timescale energy release in stellar flares.

\begin{table}[htbp]
	\centering
	\caption{Comparison of \lya\ and WL Rise Times in the 69 WLFs with Previous Studies}
	\begin{tabular}{ccccccc}
		\hline
		\hline
		Band & \multicolumn{2}{c}{This Work} & & \multicolumn{2}{c}{Previous Work} \\
		\cline{2-3} \cline{5-6}
		& Median & Mean & & Median & Mean \\
		& (min) & (min) & & (min) & (min) \\
		\midrule
		\lya\ & 2.7 & 3.9 & & 5.6\textsuperscript{a} & 7.2\textsuperscript{a} \\
		WL & 2.7 & 3.8 & & 5.9\textsuperscript{b} & 8.8\textsuperscript{b} \\
		\bottomrule
	\end{tabular}%
	\label{wllyatab2}%
	\begin{tablenotes}
	\item \textsuperscript{a} Rise times of \lya\ emission in solar flares \citep[GOES \lya\ band, see][]{Lu2021catalog}.\\
	\textsuperscript{b} Rise times of WL emission in stellar flares on solar-like stars \citep[4230--8970 \AA\ band, see][]{Yan2021MNRAS}.
	 \end{tablenotes}
\end{table}%

\subsection{Statistical Relationships of Radiated Energies}

\begin{figure}[htb]
	\centering
	\includegraphics[width=1.0\textwidth]{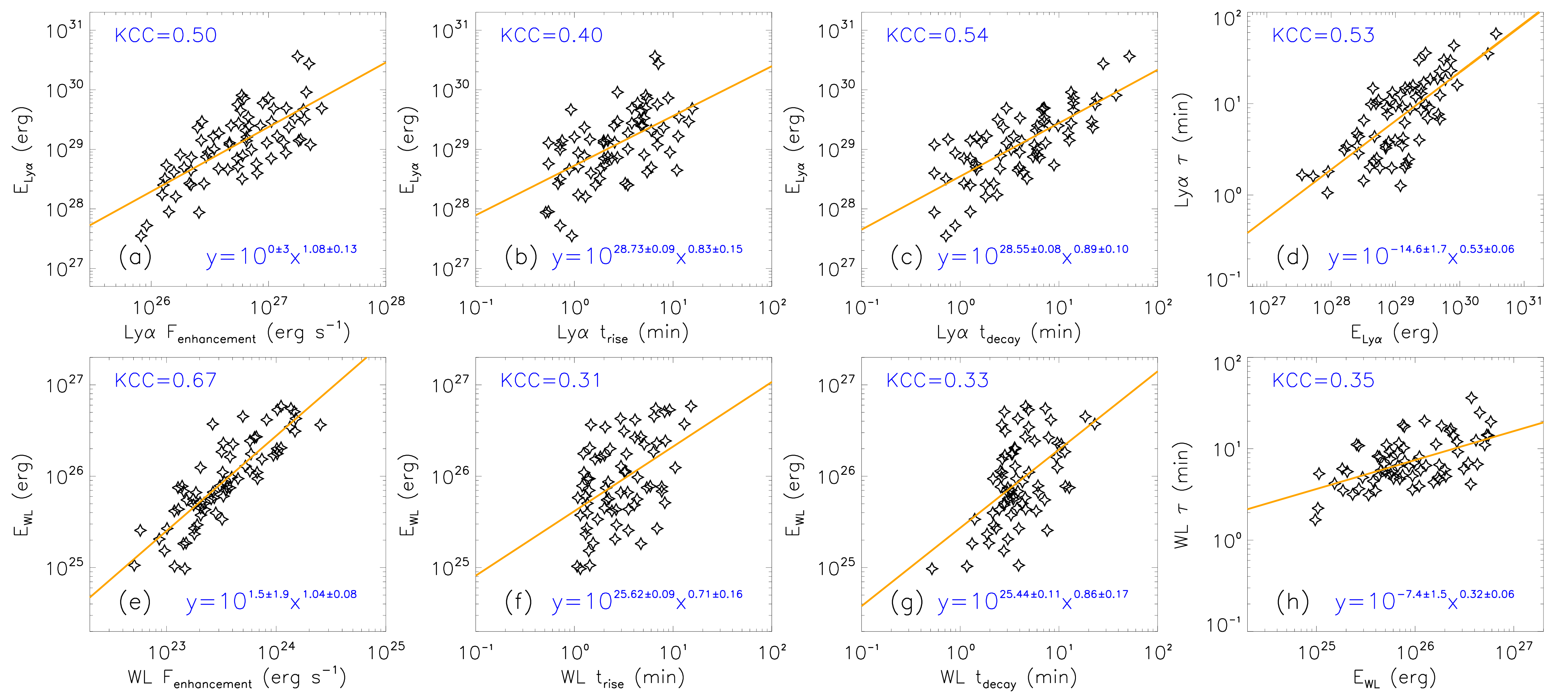}
	\caption{Relationships of the radiated energy with the peak enhancement ($F_{\text{enhancement}}$), rise time ($t_{\text{rise}}$), decay time ($t_{\text{decay}}$), and duration ($\tau$). The top and bottom rows show the results of the \lya\ and WL bands, respectively. Here the WL enhancement and energy represent the emission increase and radiated energy, respectively, in the HMI continuum near 6173~\AA\ with an assumed bandpass of $\Delta\lambda=1$~\AA. The orange solid line in each panel shows the power-law best fit, with the corresponding KCC and power-law fitting result annotated in the top-left and bottom-right corners, respectively.}
	\label{wllya_fig7}
\end{figure}

\begin{figure}[htb]
	\centering
	\includegraphics[width=1.0\textwidth]{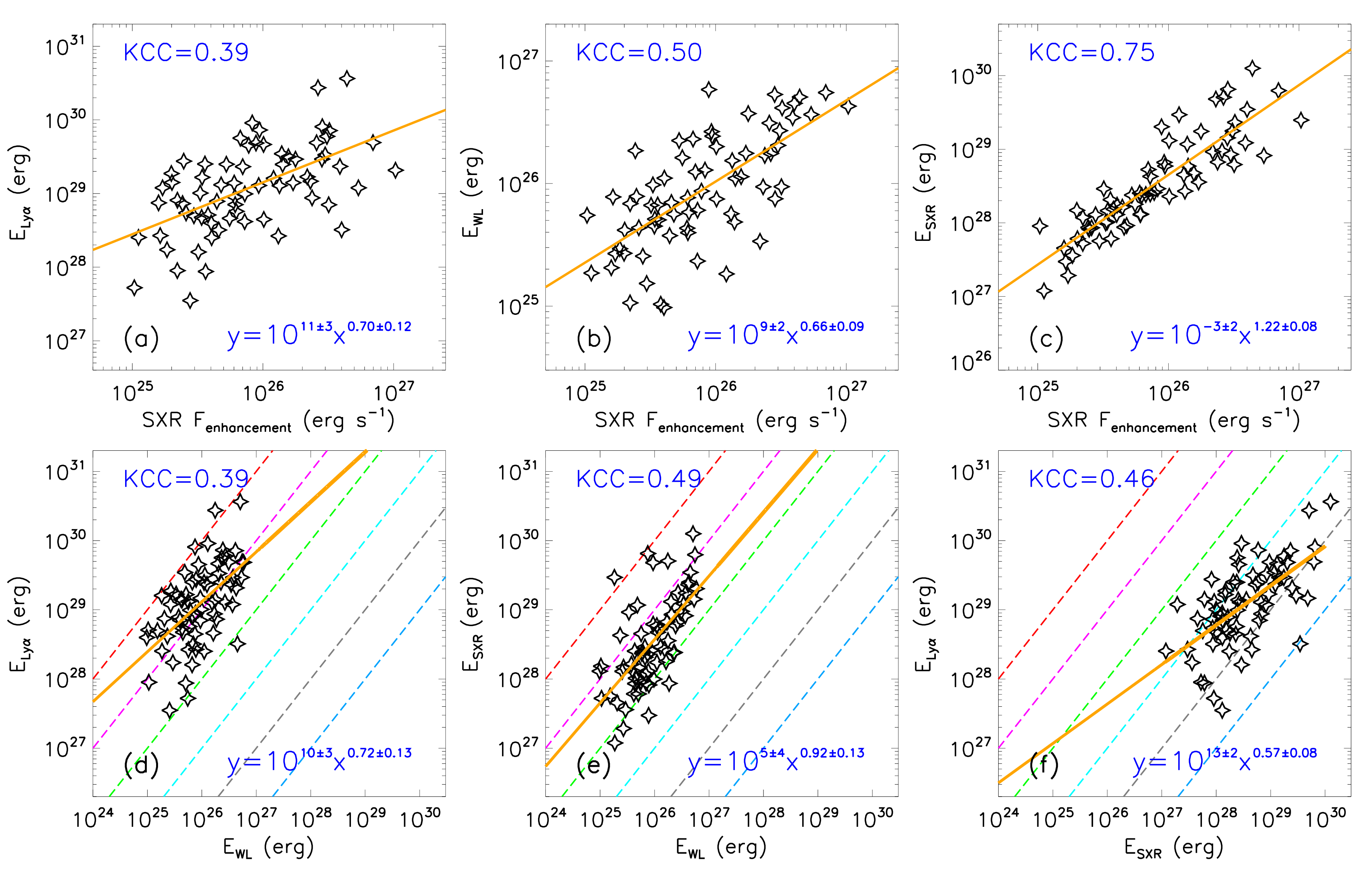}
	\caption{Relationships between the radiated energies in the \lya, WL, and SXR bands and the peak enhancement in the SXR band (top row), as well as the relationships among the energies of these three bands (bottom row). Here the WL energy represents the one in the HMI continuum near 6173~\AA\ with an assumed bandpass of $\Delta\lambda=1$~\AA. The orange solid line in each panel shows the power-law best fit, with the corresponding KCC and power-law fitting result annotated in the top-left and bottom-right corners, respectively. The red, magenta, green, cyan, gray, and sky-blue dashed lines in panels (d)--(f) indicate the energy ratios of 10000:1, 1000:1, 100:1, 10:1, 1:1, and 0.1:1, respectively.}
	\label{wllya_fig8}
\end{figure}

We investigated the statistical relationships among the radiated energies in the \lya, WL, and SXR bands, both qualitatively and quantitatively, as shown in Figures~\ref{wllya_fig7} and \ref{wllya_fig8} and Table~\ref{wllyatab3}. 
For a specific band, the radiated energy is related to its $F_{\text{enhancement}}$, $t_{\text{rise}}$, and $t_{\text{decay}}$ (or $\tau$). Figure \ref{wllya_fig7} shows the relationships between the radiated energy in the \lya\ and WL bands and the above four parameters. For the \lya\ band, $E_{\text{\lya}}$ shows positive power-law correlations with all four parameters (KCC $\ge$ 0.40), as shown in Figures \ref{wllya_fig7}(a)--(d). For the WL band (Figures \ref{wllya_fig7}(e)--(h)), $E_{\text{WL}}$ exhibits a high correlation with its $F_{\text{enhancement}}$ (KCC $= 0.67$), and somewhat moderate correlations with $t_{\text{decay}}$ (KCC $= 0.33$) and $\tau$ (KCC $= 0.35$). In particular, a positive power-law relationship exists between $E_{\text{WL}}$ and its $\tau$ (Figure \ref{wllya_fig7}(h)), with an index of 0.32$\pm$0.06, which is very close to the ideal value of 1/3 derived from the magnetic reconnection theory \citep[e.g.,][]{Shibata2011,Maehara2015,Namekata2017}.

Furthermore, we compared the relationships between the radiated energies in the WL, \lya, and SXR bands and the peak enhancement in the SXR band (i.e., SXR $F_{\text{enhancement}}$), as well as the relationships among the energies of these three bands. The corresponding scatter plots are shown in Figure \ref{wllya_fig8}. We find clear positive power-law relationships between the three energies ($E_{\text{\lya}}$, $E_{\text{WL}}$, and $E_{\text{SXR}}$) and the SXR $F_{\text{enhancement}}$ (KCCs $\ge 0.39$; Figures \ref{wllya_fig8}(a)--(c)), with the correlation between $E_{\text{WL}}$ and SXR $F_{\text{enhancement}}$ (KCC $= 0.50$; Figure \ref{wllya_fig8}(b)) stronger than that between $E_{\text{\lya}}$ and SXR $F_{\text{enhancement}}$ (KCC $= 0.39$; Figure \ref{wllya_fig8}(a)).

In addition, we find that $E_{\text{WL}}$ and $E_{\text{\lya}}$ follow a positive power-law relationship (KCC $= 0.39$; Figure~\ref{wllya_fig8}(d)). Both $E_{\text{\lya}}$ and $E_{\text{WL}}$ show even stronger positive power-law correlations with $E_{\text{SXR}}$ (KCCs $\ge 0.46$; Figures~\ref{wllya_fig8}(e) and (f)). Table~\ref{wllyatab3} lists the ranges, medians, and means of the energy ratios among the three bands. On average, $E_{\text{\lya}}$ is $\sim$2400 times $E_{\text{WL}}$ and $\sim$7 times $E_{\text{SXR}}$, while $E_{\text{SXR}}$ is $\sim$940 times $E_{\text{WL}}$.

\begin{table}[htbp]
	\centering
	\caption{Ratios of Radiated Energies in \lya, WL, and SXR bands for the 69 WLFs}
	\begin{tabular}{cccc}
		\hline
		\hline
		Energy Ratio & Range & Median & Mean \\
		\midrule
		$E_{\text{\lya}}$/$E_{\text{WL}}$ & 71.5--15588.5 & 1563.7 & 2409.0 \\
		$E_{\text{\lya}}$/$E_{\text{SXR}}$ & 0.1--62.1 & 3.9 & 7.0 \\
		$E_{\text{SXR}}$/$E_{\text{WL}}$ & 38.3--16087.3 & 345.4 & 935.4 \\
		\bottomrule
	\end{tabular}%
	\label{wllyatab3}%
\end{table}%

\section{Summary}
\label{sec:summary}

In this work, we performed a statistical analysis of the emission properties including flux enhancement, evolution timescales, and radiated energy in the \lya, WL, and SXR 1--8 \AA\ bands for 69 WLFs.
The main results are summarized as follows.

\begin{enumerate}
    \item The \lya\ contrast in these WLFs spans 0.8--28.5\% with median and mean values of 5.4\% and 7.0\%, respectively. Positive power-law correlations are found in $F_{\text{enhancement}}$ among the \lya, WL, and SXR bands (KCCs $\ge 0.49$), with the highest correlation (KCC $= 0.59$) between WL and SXR.

    \item For most events, the \lya\ peak is nearly co-temporal with the SXR time derivative, whereas the WL peak occurs co-temporally with or later than both peaks of \lya\ and the SXR derivative.

    \item The \lya\ and WL emissions show positive power-law relationships in $t_{\text{rise}}$ (KCC $= 0.41$) and $F_{\text{enhancement}}/t_{\text{rise}}$ (KCC $= 0.49$). Furthermore, their $t_{\text{rise}}$ values are very similar, typically 3--4 minutes. By contrast, these two emissions show low or medium correlations in $t_{\text{decay}}$ and $\tau$ (KCCs $\le 0.34$).

    \item $E_{\text{\lya}}$ correlates with its $F_{\text{enhancement}}$, $t_{\text{rise}}$, $t_{\text{decay}}$, and $\tau$ (KCCs $\ge 0.40$), showing the strongest relationships with $F_{\text{enhancement}}$, $t_{\text{decay}}$, and $\tau$ (KCCs $\ge 0.50$). In comparison, $E_{\text{WL}}$ correlates strongly with its $F_{\text{enhancement}}$ (KCC $= 0.67$). Additionally, $E_{\text{WL}}$ and its $\tau$ exhibit a positive power-law correlation with an index of 0.32, i.e., $\tau \propto {E}^{0.32}$.
    \item $E_{\text{\lya}}$ and $E_{\text{WL}}$ exhibit a positive power-law relationship (KCC $= 0.39$). Both energies also correlate positively with the SXR $F_{\text{enhancement}}$ (KCCs $\ge 0.39$) and $E_{\text{SXR}}$ (KCCs $\ge 0.46$). On average, $E_{\text{\lya}}$ is a few thousand of times larger than $E_{\text{WL}}$ and several times that of $E_{\text{SXR}}$. In comparison, $E_{\text{SXR}}$ is nearly a thousand times greater than $E_{\text{WL}}$.
\end{enumerate}

\section{Discussions}
\label{sec:discussion}

\subsection{\lya\ Emission Enhancement in WLFs}

For the 69 WLFs in this work, their \lya\ contrast (or relative enhancement) ranges from 0.8\% to 28.5\%, which is consistent with the results of previous statistical studies \citep[e.g.,][]{NusinovKaza2006,Milligan2020,Milligan2021}. For example, \citet{Milligan2020} statistically analyzed the contrast in the GOES \lya\ band for 477 M- and X-class flares. Their results showed that the \lya\ contrast is generally below 30\%, with a 95th percentile (P95) of $\approx$10\%. The P95 of the \lya\ contrast for the 69 WLFs in our data set is about 20\%, which is greater than that in \citet{Milligan2020}. This comparison suggests that the \lya\ contrast in WLFs, at least in a statistical sense for our sample, is stronger than that in ordinary flares. On the other hand, our statistical results show that some M- and X-class WLFs exhibit a very small \lya\ contrast (say, $<$2\%), whereas the C-class flares can produce a considerable contrast (i.e., $\sim$5\%, see Figure \ref{wllya_fig3}). Such cases have also been reported in previous studies. For example, \cite{Dominique2018} found that an X9.3 flare observed by PROBA2 has a \lya\ contrast of only 0.97\%, while \cite{Milligan2021} found a C6.6 flare observed by GOES with a \lya\ contrast as high as 7\%. These anomalous events merit a further investigation, as they are important for a comprehensive understanding of flare emission properties and energetics.

\subsection{Origin of the \lya\ Emission}
In this work, several statistical results support a nonthermal origin for the flare \lya\ emission. For instance, for most WLFs, the peak time of \lya\ light curves is nearly coincident with that of the SXR derivative curve (see Figure \ref{wllya_fig5}(b)). 
Such \lya\ emissions associated with nonthermal electrons can generally be attributed to two distinct mechanisms \citep[e.g.,][]{Hong2019,Druett2019}. 
On the one hand, when nonthermal electron beams precipitate into the lower atmospheric layer (primarily the chromosphere), hydrogen ionization and excitation rates are enhanced significantly due to strong nonthermal collisions. This process leads to an enhanced \lya\ emission. On the other hand, the subsequent rise of chromospheric temperature increases thermal collisional rates, which likewise leads to the enhancement of \lya\ emission.
Recently, an observational study by \citet{Greatorex2023} investigated the relationship between nonthermal electron properties and the \lya\ response during the impulsive phase of three M3 flares. Their results indicate that flares with harder nonthermal spectral indices tend to produce larger \lya\ contrasts. This underscores the nonthermal origin of flare \lya\ emission during the impulsive phase.

Our statistical results also suggest a thermal origin for flare \lya\ emission \citep[see also][]{Jing2020}. For example, for a small subset of WLFs, the \lya\ emission peaks later than the SXR derivative, and sometimes even later than the SXR flux peak (not shown here). This thermal origin of \lya\ emission in flares may primarily originate from heat conduction and/or the cooling of hot flare loops \citep[e.g.,][]{Milligan2016, Tian2023}. Additionally, the thermal origin of \lya\ emission can also manifest in other phenomena, such as filament eruptions \citep{RubiodaCosta2009, Wauters2022}. Radiative hydrodynamic (RHD) simulations by \citet{Hong2019} also indicate that the flare \lya\ emission can originate from both nonthermal and thermal heatings.

\subsection{Origin of the WL Emission}

The origin of the WL emission is highly complicated and remains inconclusive. Based on the spectral characteristics of WLFs, they are generally classified into two types \citep{Machado1986,Ding2007}. Type I WLFs are characterized by (1) a quasi-simultaneous peak of the WL continuum with HXR and microwave emissions, (2) enhanced and broadened hydrogen Balmer lines, and (3) the presence of a Balmer jump. In contrast, Type II WLFs lack those features and are relatively rare. Generally, Type I WLFs are believed to originate from the chromosphere, primarily due to the enhancement of the hydrogen recombination (free-bound) continuum heated by nonthermal particles (mainly electrons). Conversely, Type II WLFs are thought to originate mainly from the enhancement of the H$^{-}$ continuum emission in the photosphere, likely caused by radiative backwarming and/or direct photospheric heating. Notably, \citet{Hao2017} reported a circular-ribbon WLF in which distinct WL kernels exhibit signatures consistent with either Type I or Type II, underscoring the complexity of the heating and emission mechanisms of WLFs.

For the 69 WLFs in this study, we find both the case of quasi-simultaneous peaking (accounting for 59\%) and the case in which the WL peak lags the SXR derivative peak (38\%), as illustrated in Figure \ref{wllya_fig5}(c). 
Overall, the WL peak lags behind the SXR derivative peak with a median delay of approximately 39 s. This time delay may be attributed to hydrogen recombination processes \citep[e.g.,][]{Heinzel2014,Yyt2021RAA,Heinzel2024,Ornig2026} and/or radiative backwarming \citep[e.g.,][]{Machado1986,Ding2003,Hao2017}.
Moreover, flare-related \alfven\ waves may also transport energy from the coronal release region to the lower atmosphere and dissipate it, thereby producing the WL enhancement \citep[e.g.,][]{Emslie1982,Fletcher2008,Russell2013,Xu2025,Tian2025}. However, the WL plus other waveband light curves alone are insufficient to distinguish the types of WLFs and to disentangle the contributions of different heating mechanisms. This requires detailed case studies from observations combining with observationally-constrained RHD simulations.

\subsection{Implications for flares on Solar-type Stars}

As mentioned in Section \ref{intro}, stellar flares are often observed in WL, but their \lya\ emission is difficult to be detected due to significant interstellar absorption and scattering \citep{Linsky2014}. Our statistical results, which establish qualitative and quantitative relationships between \lya, WL, and SXR 1--8 \AA\ emissions for solar flares, serves as a critical bridge on solar-to-stellar flares.

In our work, based on calibrated data, we established positive power-law relationships for some important parameters, e.g., the flux enhancements, rise timescales, and radiated energies. We provide the fitting parameters for these scaling laws (see Figures \ref{wllya_fig4}, \ref{wllya_fig6}--\ref{wllya_fig8}) and present some quantitative ratios of these physical parameters, such as the radiated energy (Table \ref{wllyatab3}). 
Furthermore, we find $E$-$\tau$ power-law relationships in \lya\ and HMI narrow-band WL bands. For the WL band, the relationship is $\tau \propto E^{0.32 \pm 0.06}$. This result is comparable to the power-law indices of 0.38$\pm$0.06 for solar WLFs \citep{Namekata2017} and 0.39$\pm$0.03 for superflares on solar-type stars \citep{Maehara2015}, suggesting a shared underlying $E$-$\tau$ scaling. Notably, our index of 0.32 aligns more closely with the theoretical value of 1/3 predicted by simplified magnetic reconnection models. This consistency supports the universality of the magnetic reconnection mechanism across different energy scales, thereby reinforcing the validity of applying solar-derived scaling laws to interpret the energy release processes in the solar-type stars \citep[e.g.,][]{Namekata2017,Toriumi2017}.

However, we note that the solar WL parameters estimated here are derived from the HMI narrowband pseudo-continuum, which differs from the broadband WL photometry commonly used in stellar flares.
Although the different spectral bandpasses may result in variations of the continuum contrast values, the underlying physical correlations derived from spatially resolved solar data offer a valuable reference for the flares on solar-type stars.

\subsection{Impact of the Observational Temporal Resolution}

In this work, the temporal resolutions of WL, \lya, and SXR 1--8 \AA\ data are 45 s, $\sim$10 s, and $\sim$2 s, respectively, with the WL band having the lowest temporal resolution. A 45 s resolution may have a minor impact on the WL rise and decay time estimations, as these timescales are typically on the order of several minutes. However, such a resolution can certainly affect the measurements of parameters such as the peak time and enhancement in the WL band. For example, Figures~\ref{wllya_fig5}(a) and (c) show that the WL peak time is, on average, slightly delayed relative to the peaks of \lya\ and SXR time derivative by $\sim$30--40~s, although this difference lies within the relatively large timing uncertainties. In contrast, GOES has relatively higher temporal resolutions in the \lya\ and SXR bands. These data more accurately reveal the common nonthermal origin of flare \lya\ emission during the impulsive phase \citep[also see][]{Jing2020,Lu2021catalog,Greatorex2023}.

\subsection{Necessity and Recent Progress of High Cadence Observations}

Undoubtedly, in order to more accurately study the relationship between \lya\ and WL emissions in flares, higher-candence (e.g., second-scale) observations are required, especially for the WL band. High-cadence observations have revealed more fine-scale characteristics of flares. For instance, \citet{Doorsselaere2011}, using high cadence (0.05 s) \lya\ irradiance data from PROBA2/LYRA for an M2.0 flare, discovered short-period quasi-periodic pulsations (QPPs) of $\sim$8.5 s in the \lya\ band. \citet{Song2025} utilized high cadence (1--2 s) imaging observations at 3600 \AA\ continuum from the White-light Solar Telescope (WST) on the \lya\ Solar Telescope \citep[LST;][]{LH2019,Feng2019,Chen2024} onboard the Advanced Space-based Solar Observatory \citep[ASO-S;][]{Gan2019,Gan2023} to study the fine emission properties for an X-class WLF. They reported the first detection of short-period harmonic QPPs in this continuum band, with fundamental and harmonic periods of $\sim
$20 s and $\sim$11 s, respectively. 

Another instrument called the Solar Disk Imager (SDI) on ASO-S/LST can perform routine full-disk imaging observations with a cadence of $\sim$1 min as well as burst-mode observations for flare regions with a cadence of $\sim$6 s in the \lya\ band. ASO-S is also equipped with a Hard X-ray Imager \citep[HXI;][]{SuY2024}, which currently provides unique HXR observations of solar flares from the viewpoint of Earth. During flares, HXI provides high-cadence ($\sim$0.125 s) spectroscopy and imaging in the energy range of 15--300 keV. To date, ASO-S has already yielded novel results on topics such as the characteristics of \lya\ emissions in various solar activities \citep[e.g.,][]{Xue2024,LuL2024,Ying2024,LiST2024,LYL2025} and the physical properties of solar WLFs \citep[e.g.,][]{JZC2024,LiQ2024,LiY2024a,LiY2024b,Song2025,JZC2025}. For details, one can refer to \cite{Gan2025} and the references therein. In the future, based on imaging observations from ASO-S/LST, SDO/HMI, and the Extreme Ultraviolet Imager \citep[EUI;][]{EUI2020} on board SolO in \lya\ and the continua at 3600 and 6173 \AA, combined with multi-angle HXR imaging data from ASO-S/HXI and the Spectrometer Telescope for Imaging X-rays \citep[STIX;][]{STIX2020} on board SolO, we can conduct more detailed studies on the relationship between \lya\ and WL emissions in flares, as well as their physical origins and mechanisms.

\begin{acknowledgments}
We thank the anonymous referee and the statistical editor for their constructive comments and suggestions, which improved the clarity and rigor of this work.
We are grateful to the SDO and GOES teams for their open data policy. SDO is a mission of NASA's Living With a Star program. The GOES-N Series is a collaborative program between NOAA and NASA. We thank Dr. Jun Tian, Yingjie Cai, and Zhichen Jing for their helpful discussions. We also thank Drs. Yang Su and Shin Toriumi for their insightful suggestions. This work is supported by the National Key R\&D Program of China under grant 2022YFF0503004, by the Strategic Priority Research Program of the Chinese Academy of Sciences under grant XDB0560000, and by NSFC under grants 12273115 and 12233012. D.-C.S. is supported by the Jiangsu Funding Program for Excellent Postdoctoral Talent, the China Postdoctoral Science Foundation under Grant Number 2025M773193, and Natural Science Foundation of Jiangsu Province BK20241707.
\end{acknowledgments}

\bibliography{LyaWL}{}

@ARTICLE{Heinzel2024,
       author = {{Heinzel}, P.},
        title = "{Hydrogen recombination continua in stellar flares}",
      journal = {\mnras},
         year = 2024,
        month = jul,
       volume = {532},
       number = {1},
        pages = {L56-L60},
          doi = {10.1093/mnrasl/slae046},
       adsurl = {https://ui.adsabs.harvard.edu/abs/2024MNRAS.532L..56H}
}

@book{kuckartz2013statistik,
author = {Kuckartz, Udo and Rädiker, Stefan and Kollewe, Thomas and Schehl, Julia},
year = {2013},
month = {01},
pages = {},
title = {Statistik},
isbn = {978-3-531-19889-7},
doi = {10.1007/978-3-531-19890-3}
}

@ARTICLE{Russell2013,
       author = {{Russell}, A.~J.~B. and {Fletcher}, L.},
        title = "{Propagation of Alfv{\'e}nic Waves from Corona to Chromosphere and Consequences for Solar Flares}",
      journal = {\apj},
         year = 2013,
        month = mar,
       volume = {765},
       number = {2},
          eid = {81},
        pages = {81},
          doi = {10.1088/0004-637X/765/2/81},
archivePrefix = {arXiv},
       eprint = {1302.2458},
 primaryClass = {astro-ph.SR},
       adsurl = {https://ui.adsabs.harvard.edu/abs/2013ApJ...765...81R}
}

@ARTICLE{Emslie1982,
       author = {{Emslie}, A.~G. and {Sturrock}, P.~A.},
        title = "{Temperature minimum heating in solar flares by resistive dissipation of Alfv{\'e}n waves}",
      journal = {\solphys},
         year = 1982,
        month = sep,
       volume = {80},
       number = {1},
        pages = {99-112},
          doi = {10.1007/BF00153426},
       adsurl = {https://ui.adsabs.harvard.edu/abs/1982SoPh...80...99E}
}

@ARTICLE{Ornig2026,
       author = {{Ornig}, S. and {Carlsson}, M.},
        title = "{Predicted white-light solar flare emission from the F-CHROMA grid of models}",
      journal = {\aap},
         year = 2026,
        month = jan,
       volume = {705},
          eid = {A157},
        pages = {A157},
          doi = {10.1051/0004-6361/202556302},
       adsurl = {https://ui.adsabs.harvard.edu/abs/2026A&A...705A.157O}
}

@ARTICLE{Hao2017,
       author = {{Hao}, Q. and {Yang}, K. and {Cheng}, X. and {Guo}, Y. and {Fang}, C. and {Ding}, M.~D. and {Chen}, P.~F. and {Li}, Z.},
        title = "{A circular white-light flare with impulsive and gradual white-light kernels}",
      journal = {Nature Communications},
         year = 2017,
        month = dec,
       volume = {8},
          eid = {2202},
        pages = {2202},
          doi = {10.1038/s41467-017-02343-0},
archivePrefix = {arXiv},
       eprint = {1712.07279},
 primaryClass = {astro-ph.SR},
       adsurl = {https://ui.adsabs.harvard.edu/abs/2017NatCo...8.2202H}
}

@ARTICLE{Druett2019,
       author = {{Druett}, M.~K. and {Zharkova}, V.~V.},
        title = "{Non-thermal hydrogen Lyman line and continuum emission in solar flares generated by electron beams}",
      journal = {\aap},
         year = 2019,
        month = mar,
       volume = {623},
          eid = {A20},
        pages = {A20},
          doi = {10.1051/0004-6361/201732427},
       adsurl = {https://ui.adsabs.harvard.edu/abs/2019A&A...623A..20D}
}

@INPROCEEDINGS{Ding2007,
       author = {{Ding}, M.~D.},
        title = "{The Origin of Solar White-Light Flares}",
    booktitle = {The Physics of Chromospheric Plasmas},
         year = 2007,
       editor = {{Heinzel}, P. and {Dorotovi{\v{c}}}, I. and {Rutten}, R.~J.},
       series = {Astronomical Society of the Pacific Conference Series},
       volume = {368},
        month = may,
        pages = {417},
       adsurl = {https://ui.adsabs.harvard.edu/abs/2007ASPC..368..417D}
}

@INPROCEEDINGS{Tobiska2006,
       author = {{Tobiska}, W. and {Nusinov}, A.},
        title = "{ISO 21348 - Process for determining solar irradiances}",
    booktitle = {36th COSPAR Scientific Assembly},
         year = 2006,
       volume = {36},
        month = jan,
        pages = {2621},
       adsurl = {https://ui.adsabs.harvard.edu/abs/2006cosp...36.2621T}
}

@ARTICLE{Curdt2001AA,
       author = {{Curdt}, W. and {Brekke}, P. and {Feldman}, U. and {Wilhelm}, K. and {Dwivedi}, B.~N. and {Sch{\"u}hle}, U. and {Lemaire}, P.},
        title = "{The SUMER spectral atlas of solar-disk features}",
      journal = {\aap},
         year = 2001,
        month = aug,
       volume = {375},
        pages = {591-613},
          doi = {10.1051/0004-6361:20010364},
       adsurl = {https://ui.adsabs.harvard.edu/abs/2001A&A...375..591C}
}

@ARTICLE{Vernazza1981,
       author = {{Vernazza}, J.~E. and {Avrett}, E.~H. and {Loeser}, R.},
        title = "{Structure of the solar chromosphere. III. Models of the EUV brightness components of the quiet sun.}",
      journal = {\apjs},
         year = 1981,
        month = apr,
       volume = {45},
        pages = {635-725},
          doi = {10.1086/190731},
       adsurl = {https://ui.adsabs.harvard.edu/abs/1981ApJS...45..635V}
}

@ARTICLE{Henoux1995,
       author = {{Henoux}, J.~C. and {Fang}, C. and {Gan}, W.~Q.},
        title = "{Diagnostics of non-thermal processes in chromospheric flares. III. Ly{\ensuremath{\alpha}} and Ly{\ensuremath{\beta}} spectra for an atmosphere bombarded by electron or proton beams.}",
      journal = {\aap},
         year = 1995,
        month = may,
       volume = {297},
        pages = {574},
       adsurl = {https://ui.adsabs.harvard.edu/abs/1995A&A...297..574H}
}

@ARTICLE{Zhao1998,
       author = {{Zhao}, X. and {Fang}, C. and {Henoux}, J. -C.},
        title = "{Non-thermal hydrogen line emission caused by an oblique incident proton beam through charge exchange}",
      journal = {\aap},
         year = 1998,
        month = feb,
       volume = {330},
        pages = {351-358},
       adsurl = {https://ui.adsabs.harvard.edu/abs/1998A&A...330..351Z}
}

@ARTICLE{Canfield1980,
       author = {{Canfield}, R.~C. and {van Hoosier}, M.~E.},
        title = "{Observed L{\ensuremath{\alpha}} profiles for two solar flares: 14{\ensuremath{:}}12 UT 15 June, 1973 and 23{\ensuremath{:}}16 UT 21 January, 1974}",
      journal = {\solphys},
         year = 1980,
        month = aug,
       volume = {67},
       number = {2},
        pages = {339-350},
          doi = {10.1007/BF00149811},
       adsurl = {https://ui.adsabs.harvard.edu/abs/1980SoPh...67..339C}
}

@ARTICLE{Brekke1996,
       author = {{Brekke}, P. and {Rottman}, G.~J. and {Fontenla}, J. and {Judge}, P.~G.},
        title = "{The Ultraviolet Spectrum of a 3B Class Flare Observed with SOLSTICE}",
      journal = {\apj},
         year = 1996,
        month = sep,
       volume = {468},
        pages = {418},
          doi = {10.1086/177701},
       adsurl = {https://ui.adsabs.harvard.edu/abs/1996ApJ...468..418B}
}

@ARTICLE{Woods2004,
       author = {{Woods}, Thomas N. and {Eparvier}, Francis G. and {Fontenla}, Juan and {Harder}, Jerald and {Kopp}, Greg and {McClintock}, William E. and {Rottman}, Gary and {Smiley}, Byron and {Snow}, Martin},
        title = "{Solar irradiance variability during the October 2003 solar storm period}",
      journal = {\grl},
         year = 2004,
        month = may,
       volume = {31},
       number = {10},
          eid = {L10802},
        pages = {L10802},
          doi = {10.1029/2004GL019571},
       adsurl = {https://ui.adsabs.harvard.edu/abs/2004GeoRL..3110802W}
}

@ARTICLE{Allred2005,
       author = {{Allred}, Joel C. and {Hawley}, Suzanne L. and {Abbett}, William P. and {Carlsson}, Mats},
        title = "{Radiative Hydrodynamic Models of the Optical and Ultraviolet Emission from Solar Flares}",
      journal = {\apj},
         year = 2005,
        month = sep,
       volume = {630},
       number = {1},
        pages = {573-586},
          doi = {10.1086/431751},
archivePrefix = {arXiv},
       eprint = {astro-ph/0507335},
 primaryClass = {astro-ph},
       adsurl = {https://ui.adsabs.harvard.edu/abs/2005ApJ...630..573A}
}

@ARTICLE{Hong2019,
       author = {{Hong}, Jie and {Li}, Ying and {Ding}, M.~D. and {Carlsson}, Mats},
        title = "{The Response of the Ly{\ensuremath{\alpha}} Line in Different Flare Heating Models}",
      journal = {\apj},
         year = 2019,
        month = jul,
       volume = {879},
       number = {2},
          eid = {128},
        pages = {128},
          doi = {10.3847/1538-4357/ab262e},
archivePrefix = {arXiv},
       eprint = {1905.13356},
 primaryClass = {astro-ph.SR},
       adsurl = {https://ui.adsabs.harvard.edu/abs/2019ApJ...879..128H}
}

@ARTICLE{Yyt2021RAA,
       author = {{Yang}, Yu-Tong and {Hong}, Jie and {Li}, Ying and {Ding}, Ming-De and {Li}, Hui},
        title = "{Radiative hydrodynamic simulations of the spectral characteristics of solar white-light flares}",
      journal = {Research in Astronomy and Astrophysics},
         year = 2021,
        month = jan,
       volume = {21},
       number = {1},
          eid = {001},
        pages = {001},
          doi = {10.1088/1674-4527/21/1/1},
archivePrefix = {arXiv},
       eprint = {2007.08850},
 primaryClass = {astro-ph.SR},
       adsurl = {https://ui.adsabs.harvard.edu/abs/2021RAA....21....1Y}
}

@ARTICLE{Tian2022,
       author = {{Tian}, J. and {Hong}, J. and {Li}, Y. and {Ding}, M.~D.},
        title = "{An evaluation of different recipes for chromospheric radiative losses in solar flares}",
      journal = {\aap},
         year = 2022,
        month = dec,
       volume = {668},
          eid = {A96},
        pages = {A96},
          doi = {10.1051/0004-6361/202244615},
archivePrefix = {arXiv},
       eprint = {2210.04461},
 primaryClass = {astro-ph.SR},
       adsurl = {https://ui.adsabs.harvard.edu/abs/2022A&A...668A..96T}
}

@ARTICLE{Johnson2011,
       author = {{Johnson}, H. and {Raymond}, J.~C. and {Murphy}, N.~A. and {Giordano}, S. and {Ko}, Y. -K. and {Ciaravella}, A. and {Suleiman}, R.},
        title = "{Transition Region Emission from Solar Flares during the Impulsive Phase}",
      journal = {\apj},
         year = 2011,
        month = jul,
       volume = {735},
       number = {2},
          eid = {70},
        pages = {70},
          doi = {10.1088/0004-637X/735/2/70},
archivePrefix = {arXiv},
       eprint = {1104.5468},
 primaryClass = {astro-ph.SR},
       adsurl = {https://ui.adsabs.harvard.edu/abs/2011ApJ...735...70J}
}

@ARTICLE{Milligan2012,
       author = {{Milligan}, Ryan O. and {Chamberlin}, Phillip C. and {Hudson}, Hugh S. and {Woods}, Thomas N. and {Mathioudakis}, Mihalis and {Fletcher}, Lyndsay and {Kowalski}, Adam F. and {Keenan}, Francis P.},
        title = "{Observations of Enhanced Extreme Ultraviolet Continua during an X-Class Solar Flare Using SDO/EVE}",
      journal = {\apjl},
         year = 2012,
        month = mar,
       volume = {748},
       number = {1},
          eid = {L14},
        pages = {L14},
          doi = {10.1088/2041-8205/748/1/L14},
archivePrefix = {arXiv},
       eprint = {1202.1731},
 primaryClass = {astro-ph.SR},
       adsurl = {https://ui.adsabs.harvard.edu/abs/2012ApJ...748L..14M}
}

@ARTICLE{Kretzschmar2013,
       author = {{Kretzschmar}, M. and {Dominique}, M. and {Dammasch}, I.~E.},
        title = "{Sun-as-a-Star Observation of Flares in Lyman {\ensuremath{\alpha}} by the PROBA2/LYRA Radiometer}",
      journal = {\solphys},
         year = 2013,
        month = aug,
       volume = {286},
       number = {1},
        pages = {221-239},
          doi = {10.1007/s11207-012-0175-6},
archivePrefix = {arXiv},
       eprint = {1210.2169},
 primaryClass = {astro-ph.SR},
       adsurl = {https://ui.adsabs.harvard.edu/abs/2013SoPh..286..221K}
}

@ARTICLE{Milligan2016,
       author = {{Milligan}, Ryan O. and {Chamberlin}, Phillip C.},
        title = "{Anomalous temporal behaviour of broadband Ly{\ensuremath{\alpha}} observations during solar flares from SDO/EVE}",
      journal = {\aap},
         year = 2016,
        month = mar,
       volume = {587},
          eid = {A123},
        pages = {A123},
          doi = {10.1051/0004-6361/201526682},
       adsurl = {https://ui.adsabs.harvard.edu/abs/2016A&A...587A.123M}
}

@ARTICLE{Milligan2017,
       author = {{Milligan}, Ryan O. and {Fleck}, Bernhard and {Ireland}, Jack and {Fletcher}, Lyndsay and {Dennis}, Brian R.},
        title = "{Detection of Three-minute Oscillations in Full-disk Ly{\ensuremath{\alpha}} Emission during a Solar Flare}",
      journal = {\apjl},
         year = 2017,
        month = oct,
       volume = {848},
       number = {1},
          eid = {L8},
        pages = {L8},
          doi = {10.3847/2041-8213/aa8f3a},
archivePrefix = {arXiv},
       eprint = {1709.09037},
 primaryClass = {astro-ph.SR},
       adsurl = {https://ui.adsabs.harvard.edu/abs/2017ApJ...848L...8M}
}

@ARTICLE{Chamberlin2018,
       author = {{Chamberlin}, P.~C. and {Woods}, T.~N. and {Didkovsky}, L. and {Eparvier}, F.~G. and {Jones}, A.~R. and {Machol}, J.~L. and {Mason}, J.~P. and {Snow}, M. and {Thiemann}, E.~M.~B. and {Viereck}, R.~A. and {Woodraska}, D.~L.},
        title = "{Solar Ultraviolet Irradiance Observations of the Solar Flares During the Intense September 2017 Storm Period}",
      journal = {Space Weather},
         year = 2018,
        month = oct,
       volume = {16},
       number = {10},
        pages = {1470-1487},
          doi = {10.1029/2018SW001866},
       adsurl = {https://ui.adsabs.harvard.edu/abs/2018SpWea..16.1470C}
}

@ARTICLE{Dominique2018,
       author = {{Dominique}, Marie and {Zhukov}, Andrei N. and {Heinzel}, Petr and {Dammasch}, Ingolf E. and {Wauters}, Laurence and {Dolla}, Laurent and {Shestov}, Sergei and {Kretzschmar}, Matthieu and {Machol}, Janet and {Lapenta}, Giovanni and {Schmutz}, Werner},
        title = "{First Detection of Solar Flare Emission in Mid-ultraviolet Balmer Continuum}",
      journal = {\apjl},
         year = 2018,
        month = nov,
       volume = {867},
       number = {2},
          eid = {L24},
        pages = {L24},
          doi = {10.3847/2041-8213/aaeace},
archivePrefix = {arXiv},
       eprint = {1810.09835},
 primaryClass = {astro-ph.SR},
       adsurl = {https://ui.adsabs.harvard.edu/abs/2018ApJ...867L..24D}
}

@ARTICLE{LiD2020,
       author = {{Li}, Dong and {Lu}, Lei and {Ning}, Zongjun and {Feng}, Li and {Gan}, Weiqun and {Li}, Hui},
        title = "{Quasi-periodic Pulsation Detected in Ly{\ensuremath{\alpha}} Emission During Solar Flares}",
      journal = {\apj},
         year = 2020,
        month = apr,
       volume = {893},
       number = {1},
          eid = {7},
        pages = {7},
          doi = {10.3847/1538-4357/ab7cd1},
archivePrefix = {arXiv},
       eprint = {2003.01877},
 primaryClass = {astro-ph.SR},
       adsurl = {https://ui.adsabs.harvard.edu/abs/2020ApJ...893....7L}
}

@ARTICLE{Jing2020,
       author = {{Jing}, Zhichen and {Pan}, Wuqi and {Yang}, Yukun and {Song}, Dechao and {Tian}, Jun and {Li}, Y. and {Cheng}, X. and {Hong}, Jie and {Ding}, M.~D.},
        title = "{The Ly{\ensuremath{\alpha}} Emission in Solar Flares. I. A Statistical Study on Its Relationship with the 1-8 {\v{S}}Soft X-Ray Emission}",
      journal = {\apj},
         year = 2020,
        month = nov,
       volume = {904},
       number = {1},
          eid = {41},
        pages = {41},
          doi = {10.3847/1538-4357/abbacc},
archivePrefix = {arXiv},
       eprint = {2009.10358},
 primaryClass = {astro-ph.SR},
       adsurl = {https://ui.adsabs.harvard.edu/abs/2020ApJ...904...41J}
}

@ARTICLE{Lu2021catalog,
       author = {{Lu}, Lei and {Feng}, Li and {Li}, Dong and {Ying}, Beili and {Li}, Hui and {Gan}, Weiqun and {Li}, Youping and {Zhou}, Jiujiu},
        title = "{Catalog and Statistical Examinations of Ly{\ensuremath{\alpha}} Solar Flares from GOES/EUVS-E Measurements}",
      journal = {\apjs},
         year = 2021,
        month = mar,
       volume = {253},
       number = {1},
          eid = {29},
        pages = {29},
          doi = {10.3847/1538-4365/abd79b},
       adsurl = {https://ui.adsabs.harvard.edu/abs/2021ApJS..253...29L}
}

@ARTICLE{Lu2021QPP,
       author = {{Lu}, Lei and {Li}, Dong and {Ning}, Zongjun and {Feng}, Li and {Gan}, Weiqun},
        title = "{Quasi-Periodic Pulsations Detected in Ly {\ensuremath{\alpha}} and Nonthermal Emissions During Solar Flares}",
      journal = {\solphys},
         year = 2021,
        month = aug,
       volume = {296},
       number = {8},
          eid = {130},
        pages = {130},
          doi = {10.1007/s11207-021-01876-4},
archivePrefix = {arXiv},
       eprint = {2108.03820},
 primaryClass = {astro-ph.SR},
       adsurl = {https://ui.adsabs.harvard.edu/abs/2021SoPh..296..130L}
}

@ARTICLE{Milligan2021,
       author = {{Milligan}, Ryan O.},
        title = "{Solar Irradiance Variability Due to Solar Flares Observed in Lyman-Alpha Emission}",
      journal = {\solphys},
         year = 2021,
        month = mar,
       volume = {296},
       number = {3},
          eid = {51},
        pages = {51},
          doi = {10.1007/s11207-021-01796-3},
archivePrefix = {arXiv},
       eprint = {2102.00974},
 primaryClass = {astro-ph.SR},
       adsurl = {https://ui.adsabs.harvard.edu/abs/2021SoPh..296...51M}
}

@ARTICLE{Tian2023,
       author = {{Tian}, Zheng-Yuan and {Feng}, Li and {Lu}, Lei and {Xia}, Fan-Xiaoyu and {Su}, Yang and {Gan}, Wei-Qun and {Li}, Hui and {Zhou}, Yue},
        title = "{Ly{\ensuremath{\alpha}} Emission Enhancement Associated with Soft X-Ray Microflares}",
      journal = {Research in Astronomy and Astrophysics},
         year = 2023,
        month = jun,
       volume = {23},
       number = {6},
          eid = {065011},
        pages = {065011},
          doi = {10.1088/1674-4527/accc75},
       adsurl = {https://ui.adsabs.harvard.edu/abs/2023RAA....23f5011T}
}

@ARTICLE{Greatorex2023,
       author = {{Greatorex}, Harry J. and {Milligan}, Ryan O. and {Chamberlin}, Phillip C.},
        title = "{Observational Analysis of Ly{\ensuremath{\alpha}} Emission in Equivalent-magnitude Solar Flares}",
      journal = {\apj},
         year = 2023,
        month = sep,
       volume = {954},
       number = {2},
          eid = {120},
        pages = {120},
          doi = {10.3847/1538-4357/acea7f},
archivePrefix = {arXiv},
       eprint = {2306.16234},
 primaryClass = {astro-ph.SR},
       adsurl = {https://ui.adsabs.harvard.edu/abs/2023ApJ...954..120G}
}

@ARTICLE{Majury2025,
       author = {{Majury}, Luke and {Milligan}, Ryan and {Butler}, Elizabeth and {Greatorex}, Harry and {Kazachenko}, Maria},
        title = "{Spectral Irradiance Variability in Lyman-Alpha Emission During Solar Flares}",
      journal = {\solphys},
         year = 2025,
        month = may,
       volume = {300},
       number = {5},
          eid = {64},
        pages = {64},
          doi = {10.1007/s11207-025-02476-2},
archivePrefix = {arXiv},
       eprint = {2504.17667},
 primaryClass = {astro-ph.SR},
       adsurl = {https://ui.adsabs.harvard.edu/abs/2025SoPh..300...64M}
}

@ARTICLE{Evans2010,
       author = {{Evans}, J.~S. and {Strickland}, D.~J. and {Woo}, W.~K. and {McMullin}, D.~R. and {Plunkett}, S.~P. and {Viereck}, R.~A. and {Hill}, S.~M. and {Woods}, T.~N. and {Eparvier}, F.~G.},
        title = "{Early Observations by the GOES-13 Solar Extreme Ultraviolet Sensor (EUVS)}",
      journal = {\solphys},
         year = 2010,
        month = mar,
       volume = {262},
       number = {1},
        pages = {71-115},
          doi = {10.1007/s11207-009-9491-x},
       adsurl = {https://ui.adsabs.harvard.edu/abs/2010SoPh..262...71E}
}

@ARTICLE{Woods2012,
       author = {{Woods}, T.~N. and {Eparvier}, F.~G. and {Hock}, R. and {Jones}, A.~R. and {Woodraska}, D. and {Judge}, D. and {Didkovsky}, L. and {Lean}, J. and {Mariska}, J. and {Warren}, H. and {McMullin}, D. and {Chamberlin}, P. and {Berthiaume}, G. and {Bailey}, S. and {Fuller-Rowell}, T. and {Sojka}, J. and {Tobiska}, W.~K. and {Viereck}, R.},
        title = "{Extreme Ultraviolet Variability Experiment (EVE) on the Solar Dynamics Observatory (SDO): Overview of Science Objectives, Instrument Design, Data Products, and Model Developments}",
      journal = {\solphys},
         year = 2012,
        month = jan,
       volume = {275},
       number = {1-2},
        pages = {115-143},
          doi = {10.1007/s11207-009-9487-6},
       adsurl = {https://ui.adsabs.harvard.edu/abs/2012SoPh..275..115W}
}

@ARTICLE{Pesnell2012,
       author = {{Pesnell}, W. Dean and {Thompson}, B.~J. and {Chamberlin}, P.~C.},
        title = "{The Solar Dynamics Observatory (SDO)}",
      journal = {\solphys},
         year = "2012",
        month = "Jan",
       volume = {275},
       number = {1-2},
        pages = {3-15},
          doi = {10.1007/s11207-011-9841-3},
       adsurl = {https://ui.adsabs.harvard.edu/abs/2012SoPh..275....3P}
}

@ARTICLE{Dominique2013,
       author = {{Dominique}, M. and {Hochedez}, J. -F. and {Schmutz}, W. and {Dammasch}, I.~E. and {Shapiro}, A.~I. and {Kretzschmar}, M. and {Zhukov}, A.~N. and {Gillotay}, D. and {Stockman}, Y. and {BenMoussa}, A.},
        title = "{The LYRA Instrument Onboard PROBA2: Description and In-Flight Performance}",
      journal = {\solphys},
         year = 2013,
        month = aug,
       volume = {286},
       number = {1},
        pages = {21-42},
          doi = {10.1007/s11207-013-0252-5},
archivePrefix = {arXiv},
       eprint = {1302.6525},
 primaryClass = {astro-ph.IM},
       adsurl = {https://ui.adsabs.harvard.edu/abs/2013SoPh..286...21D}
}

@ARTICLE{NusinovKaza2006,
       author = {{Nusinov}, A.~A. and {Kazachevskaya}, T.~V.},
        title = "{Extreme ultraviolet and X-ray emission of solar flares as observed from the CORONAS-F spacecraft in 2001 2003}",
      journal = {Solar System Research},
         year = 2006,
        month = mar,
       volume = {40},
       number = {2},
        pages = {111-116},
          doi = {10.1134/S0038094606020043},
       adsurl = {https://ui.adsabs.harvard.edu/abs/2006SoSyR..40..111N}
}

@ARTICLE{Raulin2013,
       author = {{Raulin}, Jean-Pierre and {Trottet}, G{\'e}Rard and {Kretzschmar}, Matthieu and {Macotela}, Edith L. and {Pacini}, Alessandra and {Bertoni}, Fernando C.~P. and {Dammasch}, Ingolf E.},
        title = "{Response of the low ionosphere to X-ray and Lyman-{\ensuremath{\alpha}} solar flare emissions}",
      journal = {Journal of Geophysical Research (Space Physics)},
         year = 2013,
        month = jan,
       volume = {118},
       number = {1},
        pages = {570-575},
          doi = {10.1029/2012JA017916},
       adsurl = {https://ui.adsabs.harvard.edu/abs/2013JGRA..118..570R}
}

@ARTICLE{Milligan2020,
       author = {{Milligan}, Ryan O. and {Hudson}, Hugh S. and {Chamberlin}, Phillip C. and {Hannah}, Iain G. and {Hayes}, Laura A.},
        title = "{Lyman-alpha Variability During Solar Flares Over Solar Cycle 24 Using GOES-15/EUVS-E}",
      journal = {Space Weather},
         year = 2020,
        month = jul,
       volume = {18},
       number = {7},
          eid = {e02331},
        pages = {e02331},
          doi = {10.1029/2019SW002331},
archivePrefix = {arXiv},
       eprint = {1910.01364},
 primaryClass = {astro-ph.SR},
       adsurl = {https://ui.adsabs.harvard.edu/abs/2020SpWea..1802331M}
}

@ARTICLE{LiD2024,
       author = {{Li}, Dong and {Hong}, Zhenxiang and {Hou}, Zhenyong and {Su}, Yang},
        title = "{Localizing Quasiperiodic Pulsations in Hard X-Ray, Microwave, and Ly{\ensuremath{\alpha}} Emissions of an X6.4 Flare}",
      journal = {\apj},
         year = 2024,
        month = jul,
       volume = {970},
       number = {1},
          eid = {77},
        pages = {77},
          doi = {10.3847/1538-4357/ad566c},
archivePrefix = {arXiv},
       eprint = {2408.05463},
 primaryClass = {astro-ph.SR},
       adsurl = {https://ui.adsabs.harvard.edu/abs/2024ApJ...970...77L}
}

@ARTICLE{LiY2022,
       author = {{Li}, Y. and {Li}, Qiao and {Song}, De-Chao and {Battaglia}, Andrea Francesco and {Xiao}, Hualin and {Krucker}, S{\"a}m and {Sch{\"u}hle}, Udo and {Li}, Hui and {Gan}, Weiqun and {Ding}, M.~D.},
        title = "{The Ly{\ensuremath{\alpha}} Emission in a C1.4 Solar Flare Observed by the Extreme Ultraviolet Imager aboard Solar Orbiter}",
      journal = {\apj},
         year = 2022,
        month = sep,
       volume = {936},
       number = {2},
          eid = {142},
        pages = {142},
          doi = {10.3847/1538-4357/ac897c},
archivePrefix = {arXiv},
       eprint = {2208.06182},
 primaryClass = {astro-ph.SR},
       adsurl = {https://ui.adsabs.harvard.edu/abs/2022ApJ...936..142L}
}

@ARTICLE{SO2020,
       author = {{M{\"u}ller}, D. and {St. Cyr}, O.~C. and {Zouganelis}, I. and {Gilbert}, H.~R. and {Marsden}, R. and {Nieves-Chinchilla}, T. and {Antonucci}, E. and {Auch{\`e}re}, F. and {Berghmans}, D. and {Horbury}, T.~S. and {Howard}, R.~A. and {Krucker}, S. and {Maksimovic}, M. and {Owen}, C.~J. and {Rochus}, P. and {Rodriguez-Pacheco}, J. and {Romoli}, M. and {Solanki}, S.~K. and {Bruno}, R. and {Carlsson}, M. and {Fludra}, A. and {Harra}, L. and {Hassler}, D.~M. and {Livi}, S. and {Louarn}, P. and {Peter}, H. and {Sch{\"u}hle}, U. and {Teriaca}, L. and {del Toro Iniesta}, J.~C. and {Wimmer-Schweingruber}, R.~F. and {Marsch}, E. and {Velli}, M. and {De Groof}, A. and {Walsh}, A. and {Williams}, D.},
        title = "{The Solar Orbiter mission. Science overview}",
      journal = {\aap},
         year = 2020,
        month = oct,
       volume = {642},
          eid = {A1},
        pages = {A1},
          doi = {10.1051/0004-6361/202038467},
archivePrefix = {arXiv},
       eprint = {2009.00861},
 primaryClass = {astro-ph.SR},
       adsurl = {https://ui.adsabs.harvard.edu/abs/2020A&A...642A...1M}
}

@ARTICLE{Fletcher2011,
       author = {{Fletcher}, L. and {Dennis}, B.~R. and {Hudson}, H.~S. and {Krucker}, S. and {Phillips}, K. and {Veronig}, A. and {Battaglia}, M. and {Bone}, L. and {Caspi}, A. and {Chen}, Q. and {Gallagher}, P. and {Grigis}, P.~T. and {Ji}, H. and {Liu}, W. and {Milligan}, R.~O. and {Temmer}, M.},
        title = "{An Observational Overview of Solar Flares}",
      journal = {\ssr},
         year = 2011,
        month = sep,
       volume = {159},
       number = {1-4},
        pages = {19-106},
          doi = {10.1007/s11214-010-9701-8},
archivePrefix = {arXiv},
       eprint = {1109.5932},
 primaryClass = {astro-ph.SR},
       adsurl = {https://ui.adsabs.harvard.edu/abs/2011SSRv..159...19F}
}

@ARTICLE{Hudson2016,
       author = {{Hudson}, H.~S.},
        title = "{Chasing White-Light Flares}",
      journal = {\solphys},
         year = 2016,
        month = may,
       volume = {291},
       number = {5},
        pages = {1273-1322},
          doi = {10.1007/s11207-016-0904-3},
       adsurl = {https://ui.adsabs.harvard.edu/abs/2016SoPh..291.1273H}
}

@ARTICLE{Machado1978,
       author = {{Machado}, M.~E. and {Emslie}, A.~G. and {Brown}, J.~C.},
        title = "{The structure of the temperature minimum region in solar flares and its significance for flare heating mechanisms.}",
      journal = {\solphys},
         year = 1978,
        month = jul,
       volume = {58},
       number = {2},
        pages = {363-387},
          doi = {10.1007/BF00157282},
       adsurl = {https://ui.adsabs.harvard.edu/abs/1978SoPh...58..363M}
}

@ARTICLE{Gan1992,
       author = {{Gan}, W.~Q. and {Rieger}, E. and {Zhang}, H.~Q. and {Fang}, C.},
        title = "{The Role of Chromospheric Condensations in the Continuum Emission of White-Light Flares}",
      journal = {\apj},
         year = 1992,
        month = oct,
       volume = {397},
        pages = {694},
          doi = {10.1086/171825},
       adsurl = {https://ui.adsabs.harvard.edu/abs/1992ApJ...397..694G}
}

@ARTICLE{Fang1995AAS,
       author = {{Fang}, C. and {Ding}, M.~D.},
        title = "{On the spectral characteristics and atmospheric models of two types of white-light flares.}",
      journal = {\aaps},
         year = 1995,
        month = apr,
       volume = {110},
        pages = {99},
       adsurl = {https://ui.adsabs.harvard.edu/abs/1995A&AS..110...99F}
}

@ARTICLE{Ding2003,
       author = {{Ding}, M.~D. and {Liu}, Y. and {Yeh}, C. -T. and {Li}, J.~P.},
        title = "{Interpretation of the infrared continuum in a solar white-light flare}",
      journal = {\aap},
         year = 2003,
        month = jun,
       volume = {403},
        pages = {1151-1156},
          doi = {10.1051/0004-6361:20030428},
       adsurl = {https://ui.adsabs.harvard.edu/abs/2003A&A...403.1151D}
}

@ARTICLE{Xu2006,
       author = {{Xu}, Yan and {Cao}, Wenda and {Liu}, Chang and {Yang}, Guo and {Jing}, Ju and {Denker}, Carsten and {Emslie}, A. Gordon and {Wang}, Haimin},
        title = "{High-Resolution Observations of Multiwavelength Emissions during Two X-Class White-Light Flares}",
      journal = {\apj},
         year = 2006,
        month = apr,
       volume = {641},
       number = {2},
        pages = {1210-1216},
          doi = {10.1086/500632},
       adsurl = {https://ui.adsabs.harvard.edu/abs/2006ApJ...641.1210X}
}

@ARTICLE{Fletcher2008,
       author = {{Fletcher}, L. and {Hudson}, H.~S.},
        title = "{Impulsive Phase Flare Energy Transport by Large-Scale Alfv{\'e}n Waves and the Electron Acceleration Problem}",
      journal = {\apj},
         year = 2008,
        month = mar,
       volume = {675},
       number = {2},
        pages = {1645-1655},
          doi = {10.1086/527044},
archivePrefix = {arXiv},
       eprint = {0712.3452},
 primaryClass = {astro-ph},
       adsurl = {https://ui.adsabs.harvard.edu/abs/2008ApJ...675.1645F}
}

@ARTICLE{Heinzel2014,
       author = {{Heinzel}, P. and {Kleint}, L.},
        title = "{Hydrogen Balmer Continuum in Solar Flares Detected by the Interface Region Imaging Spectrograph (IRIS)}",
      journal = {\apjl},
         year = 2014,
        month = oct,
       volume = {794},
       number = {2},
          eid = {L23},
        pages = {L23},
          doi = {10.1088/2041-8205/794/2/L23},
archivePrefix = {arXiv},
       eprint = {1409.5680},
 primaryClass = {astro-ph.SR},
       adsurl = {https://ui.adsabs.harvard.edu/abs/2014ApJ...794L..23H}
}

@ARTICLE{Watanabe2020,
       author = {{Watanabe}, Kyoko and {Imada}, Shinsuke},
        title = "{White-light Emission and Chromospheric Response by an X1.8-class Flare on 2012 October 23}",
      journal = {\apj},
         year = 2020,
        month = mar,
       volume = {891},
       number = {1},
          eid = {88},
        pages = {88},
          doi = {10.3847/1538-4357/ab711b},
       adsurl = {https://ui.adsabs.harvard.edu/abs/2020ApJ...891...88W}
}

@ARTICLE{Xu2025,
       author = {{Xu}, Zhe and {Yan}, Xiaoli and {Li}, Zhentong and {Yang}, Liheng and {Xue}, Zhike and {Wang}, Jincheng and {Zhou}, Yian},
        title = "{High-resolution Observations of a C9.3 White-light Flare and Its Impact on the Solar Photosphere}",
      journal = {\apjl},
         year = 2025,
        month = jun,
       volume = {986},
       number = {1},
          eid = {L15},
        pages = {L15},
          doi = {10.3847/2041-8213/adddb2},
archivePrefix = {arXiv},
       eprint = {2506.08411},
 primaryClass = {astro-ph.SR},
       adsurl = {https://ui.adsabs.harvard.edu/abs/2025ApJ...986L..15X}
}

@ARTICLE{Tian2025,
       author = {{Tian}, Jun and {Li}, Ying and {Hong}, Jie and {Ding}, M.~D.},
        title = "{Continuum Enhancement and Line Asymmetries of Solar Flares in Different Heating Models: Alfv{\'e}n Waves, Nonthermal Electron Beams, and Their Joint Effect}",
      journal = {\apj},
         year = 2025,
        month = jul,
       volume = {987},
       number = {2},
          eid = {163},
        pages = {163},
          doi = {10.3847/1538-4357/addd0e},
       adsurl = {https://ui.adsabs.harvard.edu/abs/2025ApJ...987..163T}
}

@ARTICLE{LiY2024a,
       author = {{Li}, Ying and {Jing}, Zhichen and {Song}, De-Chao and {Li}, Qiao and {Tian}, Jun and {Liu}, Xiaofeng and {Wang}, Ya and {Ding}, M.~D. and {Battaglia}, Andrea Francesco and {Feng}, Li and {Li}, Hui and {Gan}, Weiqun},
        title = "{The White-light Emissions in Two X-class Flares Observed by ASO-S and CHASE}",
      journal = {\apjl},
         year = 2024,
        month = mar,
       volume = {963},
       number = {1},
          eid = {L3},
        pages = {L3},
          doi = {10.3847/2041-8213/ad27ca},
archivePrefix = {arXiv},
       eprint = {2402.07374},
 primaryClass = {astro-ph.SR},
       adsurl = {https://ui.adsabs.harvard.edu/abs/2024ApJ...963L...3L}
}

@ARTICLE{LiY2024b,
       author = {{Li}, Ying and {Liu}, Xiaofeng and {Jing}, Zhichen and {Chen}, Wei and {Li}, Qiao and {Su}, Yang and {Song}, De-Chao and {Ding}, M.~D. and {Feng}, Li and {Li}, Hui and {Gan}, Weiqun},
        title = "{Various Features of the X-class White-light Flares in Super Active Region NOAA 13664}",
      journal = {\apjl},
         year = 2024,
        month = sep,
       volume = {972},
       number = {1},
          eid = {L1},
        pages = {L1},
          doi = {10.3847/2041-8213/ad6d6c},
archivePrefix = {arXiv},
       eprint = {2408.05725},
 primaryClass = {astro-ph.SR},
       adsurl = {https://ui.adsabs.harvard.edu/abs/2024ApJ...972L...1L}
}

@ARTICLE{JZC2024,
       author = {{Jing}, Zhichen and {Li}, Ying and {Feng}, Li and {Li}, Hui and {Huang}, Yu and {Li}, Youping and {Su}, Yang and {Chen}, Wei and {Tian}, Jun and {Song}, Dechao and {Li}, Jingwei and {Xue}, Jianchao and {Zhao}, Jie and {Lu}, Lei and {Ying}, Beili and {Zhang}, Ping and {Su}, Yingna and {Zhang}, Qingmin and {Li}, Dong and {Ge}, Yunyi and {Li}, Shuting and {Li}, Qiao and {Li}, Gen and {Liu}, Xiaofeng and {Shi}, Guanglu and {Shan}, Jiahui and {Tian}, Zhengyuan and {Zhou}, Yue and {Gan}, Weiqun},
        title = "{A Statistical Study of Solar White-Light Flares Observed by the White-Light Solar Telescope of the Lyman-Alpha Solar Telescope on the Advanced Space-Based Solar Observatory (ASO-S/LST/WST) at 360 nm}",
      journal = {\solphys},
         year = 2024,
        month = feb,
       volume = {299},
       number = {2},
          eid = {11},
        pages = {11},
          doi = {10.1007/s11207-024-02251-9},
archivePrefix = {arXiv},
       eprint = {2401.07275},
 primaryClass = {astro-ph.SR},
       adsurl = {https://ui.adsabs.harvard.edu/abs/2024SoPh..299...11J}
}

@ARTICLE{JZC2025,
       author = {{Jing}, Zhichen and {Li}, Ying and {Li}, Jingwei and {Li}, Qiao},
        title = "{The M- and X-class White-light Flares in Super Active Region NOAA 13664/13697 Observed by ASO-S/LST/WST}",
      journal = {arXiv e-prints},
         year = 2025,
        month = sep,
          eid = {arXiv:2509.11029},
        pages = {arXiv:2509.11029},
          doi = {10.48550/arXiv.2509.11029},
archivePrefix = {arXiv},
       eprint = {2509.11029},
 primaryClass = {astro-ph.SR},
       adsurl = {https://ui.adsabs.harvard.edu/abs/2025arXiv250911029J}
}

@ARTICLE{Hudson2006,
       author = {{Hudson}, H.~S. and {Wolfson}, C.~J. and {Metcalf}, T.~R.},
        title = "{White-Light Flares: A TRACE/RHESSI Overview}",
      journal = {\solphys},
         year = 2006,
        month = mar,
       volume = {234},
       number = {1},
        pages = {79-93},
          doi = {10.1007/s11207-006-0056-y},
       adsurl = {https://ui.adsabs.harvard.edu/abs/2006SoPh..234...79H}
}

@ARTICLE{Song2018jj,
       author = {{Song}, Yongliang and {Tian}, Hui},
        title = "{Investigation of White-light Emission in Circular-ribbon Flares}",
      journal = {\apj},
         year = 2018,
        month = nov,
       volume = {867},
       number = {2},
          eid = {159},
        pages = {159},
          doi = {10.3847/1538-4357/aae5d1},
       adsurl = {https://ui.adsabs.harvard.edu/abs/2018ApJ...867..159S}
}

@ARTICLE{Song2018aa,
       author = {{Song}, Y.~L. and {Tian}, H. and {Zhang}, M. and {Ding}, M.~D.},
        title = "{Observations of white-light flares in NOAA active region 11515: high occurrence rate and relationship with magnetic transients}",
      journal = {\aap},
         year = 2018,
        month = jun,
       volume = {613},
          eid = {A69},
        pages = {A69},
          doi = {10.1051/0004-6361/201731817},
       adsurl = {https://ui.adsabs.harvard.edu/abs/2018A&A...613A..69S}
}

@ARTICLE{Song2020,
       author = {{Song}, Yongliang and {Tian}, Hui and {Zhu}, Xiaoshuai and {Chen}, Yajie and {Zhang}, Mei and {Zhang}, Jingwen},
        title = "{A White-light Flare Powered by Magnetic Reconnection in the Lower Solar Atmosphere}",
      journal = {\apjl},
         year = 2020,
        month = apr,
       volume = {893},
       number = {1},
          eid = {L13},
        pages = {L13},
          doi = {10.3847/2041-8213/ab83fa},
       adsurl = {https://ui.adsabs.harvard.edu/abs/2020ApJ...893L..13S}
}

@ARTICLE{LiQ2024,
       author = {{Li}, Qiao and {Li}, Ying and {Su}, Yang and {Song}, Dechao and {Li}, Hui and {Feng}, Li and {Huang}, Yu and {Li}, Youping and {Li}, Jingwei and {Zhao}, Jie and {Lu}, Lei and {Ying}, Beili and {Xue}, Jianchao and {Zhang}, Ping and {Tian}, Jun and {Liu}, Xiaofeng and {Li}, Gen and {Jing}, Zhichen and {Li}, Shuting and {Shi}, Guanglu and {Tian}, Zhengyuan and {Chen}, Wei and {Su}, Yingna and {Zhang}, Qingmin and {Li}, Dong and {Ge}, Yunyi and {Shan}, Jiahui and {Zhou}, Yue and {Lei}, Shijun and {Gan}, Weiqun},
        title = "{Spectral and Imaging Observations of a C2.3 White-Light Flare from the Advanced Space-Based Solar Observatory (ASO-S) and the Chinese H{\ensuremath{\alpha}}Solar Explorer (CHASE)}",
      journal = {\solphys},
         year = 2024,
        month = may,
       volume = {299},
       number = {5},
          eid = {73},
        pages = {73},
          doi = {10.1007/s11207-024-02313-y},
archivePrefix = {arXiv},
       eprint = {2405.01308},
 primaryClass = {astro-ph.SR},
       adsurl = {https://ui.adsabs.harvard.edu/abs/2024SoPh..299...73L}
}

@ARTICLE{Cai2024,
       author = {{Cai}, Yingjie and {Hou}, Yijun and {Li}, Ting and {Liu}, Jifeng},
        title = "{Statistics of Solar White-light Flares. I. Optimization and Application of Identification Methods}",
      journal = {\apj},
         year = 2024,
        month = nov,
       volume = {975},
       number = {1},
          eid = {69},
        pages = {69},
          doi = {10.3847/1538-4357/ad793b},
archivePrefix = {arXiv},
       eprint = {2408.05381},
 primaryClass = {astro-ph.SR},
       adsurl = {https://ui.adsabs.harvard.edu/abs/2024ApJ...975...69C}
}

@ARTICLE{Kuhar2016,
       author = {{Kuhar}, Matej and {Krucker}, S{\"a}m and {Mart{\'\i}nez Oliveros}, Juan Carlos and {Battaglia}, Marina and {Kleint}, Lucia and {Casadei}, Diego and {Hudson}, Hugh S.},
        title = "{Correlation of Hard X-Ray and White Light Emission in Solar Flares}",
      journal = {\apj},
         year = 2016,
        month = jan,
       volume = {816},
       number = {1},
          eid = {6},
        pages = {6},
          doi = {10.3847/0004-637X/816/1/6},
archivePrefix = {arXiv},
       eprint = {1511.07757},
 primaryClass = {astro-ph.SR},
       adsurl = {https://ui.adsabs.harvard.edu/abs/2016ApJ...816....6K}
}

@ARTICLE{SDC2023,
       author = {{Song}, De-Chao and {Tian}, Jun and {Li}, Y. and {Ding}, M.~D. and {Su}, Yang and {Yu}, Sijie and {Hong}, Jie and {Qiu}, Ye and {Rao}, Shihao and {Liu}, Xiaofeng and {Li}, Qiao and {Chen}, Xingyao and {Li}, Chuan and {Fang}, Cheng},
        title = "{Spectral Observations and Modeling of a Solar White-light Flare Observed by CHASE}",
      journal = {\apjl},
         year = 2023,
        month = jul,
       volume = {952},
       number = {1},
          eid = {L6},
        pages = {L6},
          doi = {10.3847/2041-8213/ace18c},
archivePrefix = {arXiv},
       eprint = {2307.12641},
 primaryClass = {astro-ph.SR},
       adsurl = {https://ui.adsabs.harvard.edu/abs/2023ApJ...952L...6S}
}

@ARTICLE{Song2025,
       author = {{Song}, De-Chao and {Dominique}, Marie and {Zimovets}, Ivan and {Li}, Qiao and {Li}, Ying and {Yu}, Fu and {Su}, Yang and {Nizamov}, B.~A. and {Wang}, Ya and {Battaglia}, Andrea Francesco and {Tian}, Jun and {Feng}, Li and {Li}, Hui and {Gan}, W.~Q.},
        title = "{Unveiling Spatiotemporal Properties of the Quasiperiodic Pulsations in the Balmer Continuum at 3600 {\r{A}} in an X-class Solar White-light Flare}",
      journal = {\apjl},
         year = 2025,
        month = apr,
       volume = {983},
       number = {2},
          eid = {L41},
        pages = {L41},
          doi = {10.3847/2041-8213/adc4e9},
archivePrefix = {arXiv},
       eprint = {2504.02415},
 primaryClass = {astro-ph.SR},
       adsurl = {https://ui.adsabs.harvard.edu/abs/2025ApJ...983L..41S}
}

@ARTICLE{Battaglia2025,
       author = {{Battaglia}, Andrea Francesco and {Krucker}, S{\"a}m},
        title = "{New insights into the proton precipitation sites in solar flares}",
      journal = {\aap},
         year = 2025,
        month = feb,
       volume = {694},
          eid = {A58},
        pages = {A58},
          doi = {10.1051/0004-6361/202453144},
archivePrefix = {arXiv},
       eprint = {2412.11490},
 primaryClass = {astro-ph.SR},
       adsurl = {https://ui.adsabs.harvard.edu/abs/2025A&A...694A..58B}
}

@ARTICLE{LiD2025,
       author = {{Li}, Dong},
        title = "{Localizing short-period pulsations in hard X-rays and {\ensuremath{\gamma}}-rays during an X9.0 flare}",
      journal = {\aap},
         year = 2025,
        month = mar,
       volume = {695},
          eid = {L4},
        pages = {L4},
          doi = {10.1051/0004-6361/202453613},
archivePrefix = {arXiv},
       eprint = {2502.14262},
 primaryClass = {astro-ph.SR},
       adsurl = {https://ui.adsabs.harvard.edu/abs/2025A&A...695L...4L}
}

@ARTICLE{Linsky2014,
       author = {{Linsky}, Jeffrey L. and {Fontenla}, Juan and {France}, Kevin},
        title = "{The Intrinsic Extreme Ultraviolet Fluxes of F5 V TO M5 V Stars}",
      journal = {\apj},
         year = 2014,
        month = jan,
       volume = {780},
       number = {1},
          eid = {61},
        pages = {61},
          doi = {10.1088/0004-637X/780/1/61},
archivePrefix = {arXiv},
       eprint = {1310.1360},
 primaryClass = {astro-ph.SR},
       adsurl = {https://ui.adsabs.harvard.edu/abs/2014ApJ...780...61L}
}

@INPROCEEDINGS{Hanser1996,
       author = {{Hanser}, Frederick A. and {Sellers}, Francis B.},
        title = "{Design and calibration of the GOES-8 solar x-ray sensor: the XRS}",
    booktitle = {GOES-8 and Beyond},
         year = 1996,
       editor = {{Washwell}, Edward R.},
       series = {Society of Photo-Optical Instrumentation Engineers (SPIE) Conference Series},
       volume = {2812},
        month = oct,
        pages = {344-352},
          doi = {10.1117/12.254082},
       adsurl = {https://ui.adsabs.harvard.edu/abs/1996SPIE.2812..344H}
}

@ARTICLE{Scherrer2012,
       author = {{Scherrer}, P.~H. and {Schou}, J. and {Bush}, R.~I. and {Kosovichev}, A.~G. and {Bogart}, R.~S. and {Hoeksema}, J.~T. and {Liu}, Y. and {Duvall}, T.~L. and {Zhao}, J. and {Title}, A.~M. and {Schrijver}, C.~J. and {Tarbell}, T.~D. and {Tomczyk}, S.},
        title = "{The Helioseismic and Magnetic Imager (HMI) Investigation for the Solar Dynamics Observatory (SDO)}",
      journal = {\solphys},
         year = 2012,
        month = jan,
       volume = {275},
       number = {1-2},
        pages = {207-227},
          doi = {10.1007/s11207-011-9834-2},
       adsurl = {https://ui.adsabs.harvard.edu/abs/2012SoPh..275..207S}
}

@ARTICLE{SSW1998,
       author = {{Freeland}, S.~L. and {Handy}, B.~N.},
        title = "{Data Analysis with the SolarSoft System}",
      journal = {\solphys},
         year = 1998,
        month = oct,
       volume = {182},
       number = {2},
        pages = {497-500},
          doi = {10.1023/A:1005038224881},
       adsurl = {https://ui.adsabs.harvard.edu/abs/1998SoPh..182..497F}
}

@ARTICLE{Buitrago-Casas2015,
       author = {{Buitrago-Casas}, J.~C. and {Mart{\'\i}nez Oliveros}, J.~C. and {Lindsey}, C. and {Calvo-Mozo}, B. and {Krucker}, S. and {Glesener}, L. and {Zharkov}, S.},
        title = "{A Statistical Correlation of Sunquakes Based on Their Seismic and White-Light Emission}",
      journal = {\solphys},
         year = 2015,
        month = nov,
       volume = {290},
       number = {11},
        pages = {3151-3162},
          doi = {10.1007/s11207-015-0786-9},
archivePrefix = {arXiv},
       eprint = {1502.07798},
 primaryClass = {astro-ph.SR},
       adsurl = {https://ui.adsabs.harvard.edu/abs/2015SoPh..290.3151B}
}

@ARTICLE{Huang2016,
       author = {{Huang}, Neng-Yi and {Xu}, Yan and {Wang}, Haimin},
        title = "{The Energetics of White-light Flares Observed by SDO/HMI and RHESSI}",
      journal = {Research in Astronomy and Astrophysics},
         year = 2016,
        month = nov,
       volume = {16},
       number = {11},
          eid = {177},
        pages = {177},
          doi = {10.1088/1674-4527/16/11/177},
archivePrefix = {arXiv},
       eprint = {1608.06015},
 primaryClass = {astro-ph.SR},
       adsurl = {https://ui.adsabs.harvard.edu/abs/2016RAA....16..177H}
}

@ARTICLE{Namekata2017,
       author = {{Namekata}, Kosuke and {Sakaue}, Takahito and {Watanabe}, Kyoko and {Asai}, Ayumi and {Maehara}, Hiroyuki and {Notsu}, Yuta and {Notsu}, Shota and {Honda}, Satoshi and {Ishii}, Takako T. and {Ikuta}, Kai and {Nogami}, Daisaku and {Shibata}, Kazunari},
        title = "{Statistical Studies of Solar White-light Flares and Comparisons with Superflares on Solar-type Stars}",
      journal = {\apj},
         year = 2017,
        month = dec,
       volume = {851},
       number = {2},
          eid = {91},
        pages = {91},
          doi = {10.3847/1538-4357/aa9b34},
archivePrefix = {arXiv},
       eprint = {1710.11325},
 primaryClass = {astro-ph.SR},
       adsurl = {https://ui.adsabs.harvard.edu/abs/2017ApJ...851...91N}
}

@ARTICLE{Watanabe2017,
       author = {{Watanabe}, Kyoko and {Kitagawa}, Jun and {Masuda}, Satoshi},
        title = "{Characteristics that Produce White-light Enhancements in Solar Flares Observed by Hinode/SOT}",
      journal = {\apj},
         year = 2017,
        month = dec,
       volume = {850},
       number = {2},
          eid = {204},
        pages = {204},
          doi = {10.3847/1538-4357/aa9659},
archivePrefix = {arXiv},
       eprint = {1710.09531},
 primaryClass = {astro-ph.SR},
       adsurl = {https://ui.adsabs.harvard.edu/abs/2017ApJ...850..204W}
}

@ARTICLE{Song2018ApJb,
       author = {{Song}, Y.~L. and {Guo}, Y. and {Tian}, H. and {Zhu}, X.~S. and {Zhang}, M. and {Zhu}, Y.~J.},
        title = "{Observations of a White-light Flare Associated with a Filament Eruption}",
      journal = {\apj},
         year = 2018,
        month = feb,
       volume = {854},
       number = {1},
          eid = {64},
        pages = {64},
          doi = {10.3847/1538-4357/aaa7f1},
archivePrefix = {arXiv},
       eprint = {1801.04408},
 primaryClass = {astro-ph.SR},
       adsurl = {https://ui.adsabs.harvard.edu/abs/2018ApJ...854...64S}
}

@ARTICLE{Lee2017,
       author = {{Lee}, Kyoung-Sun and {Imada}, Shinsuke and {Watanabe}, Kyoko and {Bamba}, Yumi and {Brooks}, David H.},
        title = "{IRIS, Hinode, SDO, and RHESSI Observations of a White Light Flare Produced Directly by Nonthermal Electrons}",
      journal = {\apj},
         year = 2017,
        month = feb,
       volume = {836},
       number = {2},
          eid = {150},
        pages = {150},
          doi = {10.3847/1538-4357/aa5b8b},
archivePrefix = {arXiv},
       eprint = {1701.06286},
 primaryClass = {astro-ph.SR},
       adsurl = {https://ui.adsabs.harvard.edu/abs/2017ApJ...836..150L}
}

@ARTICLE{Scherrer1995,
       author = {{Scherrer}, P.~H. and {Bogart}, R.~S. and {Bush}, R.~I. and {Hoeksema}, J.~T. and {Kosovichev}, A.~G. and {Schou}, J. and {Rosenberg}, W. and {Springer}, L. and {Tarbell}, T.~D. and {Title}, A. and {Wolfson}, C.~J. and {Zayer}, I. and {MDI Engineering Team}},
        title = "{The Solar Oscillations Investigation - Michelson Doppler Imager}",
      journal = {\solphys},
         year = 1995,
        month = dec,
       volume = {162},
       number = {1-2},
        pages = {129-188},
          doi = {10.1007/BF00733429},
       adsurl = {https://ui.adsabs.harvard.edu/abs/1995SoPh..162..129S}
}

@ARTICLE{SOHO1995,
       author = {{Domingo}, V. and {Fleck}, B. and {Poland}, A.~I.},
        title = "{The SOHO Mission: an Overview}",
      journal = {\solphys},
         year = 1995,
        month = dec,
       volume = {162},
       number = {1-2},
        pages = {1-37},
          doi = {10.1007/BF00733425},
       adsurl = {https://ui.adsabs.harvard.edu/abs/1995SoPh..162....1D}
}

@ARTICLE{Baliukin2019,
       author = {{Baliukin}, I.~I. and {Bertaux}, J. -L. and {Qu{\'e}merais}, E. and {Izmodenov}, V.~V. and {Schmidt}, W.},
        title = "{SWAN/SOHO Lyman-{\ensuremath{\alpha}} Mapping: The Hydrogen Geocorona Extends Well Beyond the Moon}",
      journal = {Journal of Geophysical Research (Space Physics)},
         year = 2019,
        month = feb,
       volume = {124},
       number = {2},
        pages = {861-885},
          doi = {10.1029/2018JA026136},
       adsurl = {https://ui.adsabs.harvard.edu/abs/2019JGRA..124..861B}
}

@ARTICLE{Joshi2019,
       author = {{Joshi}, Anita and {Chandra}, Ramesh},
        title = "{North-South Distribution and Asymmetry of GOES SXR Flares during Solar Cycle 24}",
      journal = {Open Astronomy},
         year = 2019,
        month = dec,
       volume = {28},
       number = {1},
        pages = {228-235},
          doi = {10.1515/astro-2019-0019},
archivePrefix = {arXiv},
       eprint = {2007.06998},
 primaryClass = {astro-ph.SR},
       adsurl = {https://ui.adsabs.harvard.edu/abs/2019OAst...28..228J}
}

@ARTICLE{Maehara2015,
       author = {{Maehara}, Hiroyuki and {Shibayama}, Takuya and {Notsu}, Yuta and {Notsu}, Shota and {Honda}, Satoshi and {Nogami}, Daisaku and {Shibata}, Kazunari},
        title = "{Statistical properties of superflares on solar-type stars based on 1-min cadence data}",
      journal = {Earth, Planets and Space},
         year = 2015,
        month = dec,
       volume = {67},
          eid = {59},
        pages = {59},
          doi = {10.1186/s40623-015-0217-z},
archivePrefix = {arXiv},
       eprint = {1504.00074},
 primaryClass = {astro-ph.SR},
       adsurl = {https://ui.adsabs.harvard.edu/abs/2015EP&S...67...59M}
}

@ARTICLE{Kowalski2024,
       author = {{Kowalski}, Adam F.},
        title = "{Stellar flares}",
      journal = {Living Reviews in Solar Physics},
         year = 2024,
        month = dec,
       volume = {21},
       number = {1},
          eid = {1},
        pages = {1},
          doi = {10.1007/s41116-024-00039-4},
archivePrefix = {arXiv},
       eprint = {2402.07885},
 primaryClass = {astro-ph.SR},
       adsurl = {https://ui.adsabs.harvard.edu/abs/2024LRSP...21....1K}
}

@ARTICLE{Greatorex2024,
       author = {{Greatorex}, Harry J. and {Milligan}, Ryan O. and {Dammasch}, Ingolf E.},
        title = "{On the Instrumental Discrepancies in Lyman-Alpha Observations of Solar Flares}",
      journal = {\solphys},
         year = 2024,
        month = nov,
       volume = {299},
       number = {11},
          eid = {162},
        pages = {162},
          doi = {10.1007/s11207-024-02407-7},
archivePrefix = {arXiv},
       eprint = {2411.00736},
 primaryClass = {astro-ph.SR},
       adsurl = {https://ui.adsabs.harvard.edu/abs/2024SoPh..299..162G}
}

@ARTICLE{Kleint2016,
       author = {{Kleint}, Lucia and {Heinzel}, Petr and {Judge}, Phil and {Krucker}, S{\"a}m},
        title = "{Continuum Enhancements in the Ultraviolet, the Visible and the Infrared during the X1 Flare on 2014 March 29}",
      journal = {\apj},
         year = 2016,
        month = jan,
       volume = {816},
       number = {2},
          eid = {88},
        pages = {88},
          doi = {10.3847/0004-637X/816/2/88},
archivePrefix = {arXiv},
       eprint = {1511.04161},
 primaryClass = {astro-ph.SR},
       adsurl = {https://ui.adsabs.harvard.edu/abs/2016ApJ...816...88K}
}

@ARTICLE{Heinzel2017,
       author = {{Heinzel}, P. and {Kleint}, L. and {Ka{\v{s}}parov{\'a}}, J. and {Krucker}, S.},
        title = "{On the Nature of Off-limb Flare Continuum Sources Detected by SDO/HMI}",
      journal = {\apj},
         year = 2017,
        month = sep,
       volume = {847},
       number = {1},
          eid = {48},
        pages = {48},
          doi = {10.3847/1538-4357/aa86ef},
archivePrefix = {arXiv},
       eprint = {1709.06377},
 primaryClass = {astro-ph.SR},
       adsurl = {https://ui.adsabs.harvard.edu/abs/2017ApJ...847...48H}
}

@ARTICLE{jejcic18,
       author = {{Jej{\v{c}}i{\v{c}}}, S. and {Kleint}, L. and {Heinzel}, P.},
        title = "{High-density Off-limb Flare Loops Observed by SDO}",
      journal = {\apj},
         year = 2018,
        month = nov,
       volume = {867},
       number = {2},
          eid = {134},
        pages = {134},
          doi = {10.3847/1538-4357/aae650},
archivePrefix = {arXiv},
       eprint = {1810.02431},
 primaryClass = {astro-ph.SR},
       adsurl = {https://ui.adsabs.harvard.edu/abs/2018ApJ...867..134J}
}

@ARTICLE{Castellanos2020,
       author = {{Castellanos Dur{\'a}n}, J. Sebasti{\'a}n and {Kleint}, Lucia},
        title = "{The Statistical Relationship between White-light Emission and Photospheric Magnetic Field Changes in Flares}",
      journal = {\apj},
         year = 2020,
        month = dec,
       volume = {904},
       number = {2},
          eid = {96},
        pages = {96},
          doi = {10.3847/1538-4357/ab9c1e},
archivePrefix = {arXiv},
       eprint = {2007.02954},
 primaryClass = {astro-ph.SR},
       adsurl = {https://ui.adsabs.harvard.edu/abs/2020ApJ...904...96C}
}

@ARTICLE{GarciaRivas2024,
       author = {{Garc{\'\i}a-Rivas}, M. and {Ka{\v{s}}parov{\'a}}, J. and {Berlicki}, A. and {{\v{S}}vanda}, M. and {Dud{\'\i}k}, J. and {{\v{C}}tvrte{\v{c}}ka}, D. and {Zapi{\'o}r}, M. and {Liu}, W. and {Sobotka}, M. and {Pavelkov{\'a}}, M. and {Motorina}, G.~G.},
        title = "{Flare heating of the chromosphere: Observations of flare continuum from GREGOR and IRIS}",
      journal = {\aap},
         year = 2024,
        month = oct,
       volume = {690},
          eid = {A254},
        pages = {A254},
          doi = {10.1051/0004-6361/202451219},
archivePrefix = {arXiv},
       eprint = {2408.10813},
 primaryClass = {astro-ph.SR},
       adsurl = {https://ui.adsabs.harvard.edu/abs/2024A&A...690A.254G}
}

@INPROCEEDINGS{Neckel1994,
       author = {{Neckel}, H.},
        title = "{Solar Absolute Reference Spectrum}",
    booktitle = {Invited Papers from IAU Colloquium 143: The Sun as a Variable Star: Solar and Stellar Irradiance Variations},
         year = 1994,
       editor = {{Pap}, J.~M. and {Frohlich}, C. and {Hudson}, H.~S. and {Solanki}, S.~K.},
        month = jan,
        pages = {37},
       adsurl = {https://ui.adsabs.harvard.edu/abs/1994svsp.coll...37N}
}

@ARTICLE{Neupert1968,
       author = {{Neupert}, Werner M.},
        title = "{Comparison of Solar X-Ray Line Emission with Microwave Emission during Flares}",
      journal = {\apjl},
         year = 1968,
        month = jul,
       volume = {153},
        pages = {L59},
          doi = {10.1086/180220},
       adsurl = {https://ui.adsabs.harvard.edu/abs/1968ApJ...153L..59N}
}

@ARTICLE{Yan2021MNRAS,
       author = {{Yan}, Y. and {He}, H. and {Li}, C. and {Esamdin}, A. and {Tan}, B.~L. and {Zhang}, L.~Y. and {Wang}, H.},
        title = "{Characteristic time of stellar flares on Sun-like stars}",
      journal = {\mnras},
         year = 2021,
        month = jul,
       volume = {505},
       number = {1},
        pages = {L79-L83},
          doi = {10.1093/mnrasl/slab055},
archivePrefix = {arXiv},
       eprint = {2105.12375},
 primaryClass = {astro-ph.SR},
       adsurl = {https://ui.adsabs.harvard.edu/abs/2021MNRAS.505L..79Y}
}
\bibliographystyle{aasjournal}

\end{document}